\documentclass[reprint,aps,prd,onecolumn,notitlepage,showpacs,nofootinbib,preprintnumbers,superscriptaddress]{revtex4-1}
\usepackage{amsmath}
\usepackage{amsfonts}
\usepackage{amssymb}
\usepackage{latexsym}
\usepackage{color}
\usepackage{graphicx}
\usepackage{bm}
\usepackage{epsfig}
\usepackage[english]{babel}
\usepackage{hyperref}
\usepackage{times}
\usepackage{comment}
\usepackage{multirow}

\def\iso{\mathrm{iso}}
\def\diag{\mathrm{diag}}

\begin{document}

\title{Primordial perturbations generated by Higgs field and $R^2$ operator}
\author{Yun-Chao Wang}
\author{Tower Wang}
\email[Electronic address:]{twang@phy.ecnu.edu.cn}
\affiliation{Department of Physics, East China Normal University,\\
Shanghai 200241, China\\ \vspace{0.2cm}}
\date{\today\\ \vspace{1cm}}
\begin{abstract}
If the very early Universe is dominated by the non-minimally coupled Higgs field and Starobinsky's curvature-squared term together, the potential diagram would mimic the landscape of a valley, serving as a cosmological attractor. The inflationary dynamics along this valley is studied, model parameters are constrained against observational data, and the effect of isocurvature perturbation is estimated.
\end{abstract}


\maketitle




\section{Introduction}\label{sect-intro}

Recent observations of the cosmic microwave background (CMB) radiation \cite{Ade:2015tva,Ade:2015xua,Ade:2015lrj} have acquired more information about the very early Universe. They focused our attention on a smaller region in the ($n_s$, $r$) plane, where $n_s$ is the scalar spectral index and $r$ is the tensor-to-scalar ratio, thus giving us more clues to the standard model of slow-roll single-field inflation. Meanwhile, the data also shrunk the parameter space of signatures of new physics, such as the power spectral features, the primordial non-Gaussianities, the residual isocurvature modes, the power asymmetry and the running of spectral index, therefore placing stringent constraints on more complicated inflationary models.

Among typical single-field models of slow-roll inflation, two thrifty models became favored by data and popular recently: the Starobinsky's $R^2$ inflation \cite{Starobinsky:1980te,Starobinsky:1983zz} and the non-minimally coupled Higgs inflation \cite{Bezrukov:2007ep}. In Ref. \cite{Starobinsky:1980te}, the $R^2$ cosmological model was proposed to get a nonsingular isotropic homogeneous picture of the Universe. As a solution to the horizon and flatness problems, the inflationary universe scenario \cite{Guth:1980zm,Linde:1981mu,Albrecht:1982wi} is realized often by a scalar field. Both of them are able to generate a perturbation spectrum which may lead to the microwave background anisotropy and the structure formation in the Universe. Interestingly, in the absence of matter, the Starobinsky model is conformally equivalent to the Einstein gravity plus a scalar field \cite{Whitt:1984pd}. In the same case, it is also equivalent to the large-field model of non-minimally coupled Higgs inflation \cite{Bezrukov:2007ep,Kehagias:2013mya}. Nevertheless, the $R^2$ inflation and the Higgs inflation can be distinguished observationally by considering their post-inflationary interactions with matter fields \cite{Bezrukov:2011gp}.

When both a scalar field and a curvature-squared term are effective in driving inflation, inflationary dynamics will be more complicated. To the best of our knowledge, the exploration of such models dates back to three decades ago \cite{Kofman:1985aw,Starobinsky:1986fxa,Muller:1988db,Gottlober:1990um}. For instance, Ref. \cite{Gottlober:1990um} analyzed a specific model of this type where the scalar potential is of the quadratic form. It was found that, in certain conditions, the model can generate two consecutive inflationary stages and a break in the perturbation spectrum. In recent years, there are reviving interests in this class of models, with attention turned to the case of a single inflationary phase, see Refs. \cite{Cardenas:2003tg,vandeBruck:2015xpa,Bamba:2015uxa,Bamba:2016fzw,Kaneda:2015jma} as a partial list. The model of Ref. \cite{Gottlober:1990um} was revisited in Ref. \cite{Cardenas:2003tg} under the approximation (10) therein, and in Ref. \cite{vandeBruck:2015xpa} to the second order in slow-roll parameters. In Refs. \cite{Bamba:2015uxa,Bamba:2016fzw}, inflationary cosmology has been explored in a theory of two scalar fields non-minimally coupled to the Ricci scalar with an extra $R^2$ term. In Ref. \cite{Kaneda:2015jma}, the special case with $V_H=0$ was studied analytically and numerically.

In contrast, if one assumes that the Higgs field takes a small value in the very early Universe, then the $R^2$ term is effective in driving inflation. For this case, it was argued in Ref. \cite{Torabian:2014nva} that a non-minimal coupling of Higgs field to curvature could alleviate fine-tuning problems of the initial Higgs value, and it was shown in Ref. \cite{Calmet:2016fsr} that the Higgs boson can create a large Wilson coefficient for the $R^2$ operator.

In this paper, we will study the inflationary dynamics driven by the $R^2$ term together with a non-minimally coupled Higgs field. In Sec. \ref{sect-bview}, we will write this model in the Einstein frame and illustrate the valley-like potential landscape. In Sec. \ref{sect-1f}, slow-roll inflation in the valley will be analyzed with a single-field approximation. In Sec. \ref{sect-data}, we will fit the model parameters with Planck 2015 data, and then demonstrate the precision of cosmological parameters with best-fit models. In Sec. \ref{sect-2f}, we will perform a two-field numerical simulation to check the single-field approximation. A few subtle points of this model will be discussed in Sec. \ref{sect-disc}. In appendix \ref{app-vall}, we will develop an analytical method to locate the valley of a two-field potential. Appendix \ref{app-miso} will present some analytical formulae for the mass of isocurvature mode and the slow-turn parameter, which are useful for the numerical simulation in Sec. \ref{sect-2f}.

Throughout this paper, $M_p=(8\pi G)^{-1/2}$ is the reduced Plank mass. The derivatives of potential are denoted by $V_{\phi}=dV/d\phi$, $V_{\chi}=dV/d\chi$, $V_{\phi\phi}=d^2V/d\phi^2$, etc. The metric and the Ricci tensor are denoted by $g_{\mu\nu}$, $R_{\mu\nu}$ in the Jordan frame, and $\tilde{g}_{\mu\nu}$, $\tilde{R}_{\mu\nu}$ in the Einstein frame. The quantities with a subscript star are evaluated at Hubble crossing $k=a_{\ast}H_{\ast}$, and the quantities with a subscript $e$ are evaluated at the end of inflation. In Sec. \ref{sect-1f}, we will introduce the notation $\Lambda$ in Eq. \eqref{Lambda}, and  $\rho=M_p^2/(\xi\chi^2)$. The readers should not confuse the \emph{true} e-folding number $N=\ln(a_e/a)$ with the \emph{uncalibrated} e-folding number $\mathcal{N}$ defined by Eq. \eqref{unefolds}.

\section{Bird view}\label{sect-bview}
The full action of Higgs inflation model \cite{Bezrukov:2007ep} augmented with an $R^2$ term can be written as
\begin{eqnarray}\label{act}
\nonumber\mathcal{S}&=&\int d^4x\sqrt{-g}\left[\frac{M_p^2}{2}R+\frac{1}{2}\xi\chi^2R+\frac{M_p^2}{12M^2}R^2-\frac{1}{2}g^{\mu\nu}\partial_{\mu}\chi\partial_{\nu}\chi-\frac{1}{4}\lambda\left(\chi^2-v^2\right)^2\right]\\
&=&\int d^4x\sqrt{-\tilde{g}}\left[\frac{M_p^2}{2}\tilde{R}-\frac{1}{2}\tilde{g}^{\mu\nu}\partial_{\mu}\phi\partial_{\nu}\phi-\frac{1}{2}e^{-\sqrt{\frac{2}{3}}\phi/M_p}\tilde{g}^{\mu\nu}\partial_{\mu}\chi\partial_{\nu}\chi-V(\phi,\chi)\right].
\end{eqnarray}
As an important example, one can have in mind that $\chi$ is the Higgs boson in the standard model of particle physics \cite{Fumagalli:2016lls,Fumagalli:2016sof,Lalak:2016idb}, but our discussion will be quite general for scalars with a quartic potential term. From now on, we assume $v^2\ll\chi^2$ during inflation and thus set the parameter $v^2$ to zero. Then the potential in the Einstein frame is
\begin{equation}\label{V}
V(\phi,\chi)=\frac{3}{4}M_p^2M^2e^{-2\sqrt{\frac{2}{3}}\phi/M_p}\left(e^{\sqrt{\frac{2}{3}}\phi/M_p}-1-\frac{1}{M_p^2}\xi\chi^2\right)^2+\frac{1}{4}\lambda\chi^4e^{-2\sqrt{\frac{2}{3}}\phi/M_p}.
\end{equation}
The Jordan and Einstein frames are related by the conformal transformation $\tilde{g}_{\mu\nu}=Fg_{\mu\nu}$ with
\begin{equation}\label{F}
F=1+\frac{1}{M_p^2}\xi\chi^2+\frac{1}{3M^2}R=e^{\sqrt{\frac{2}{3}}\phi/M_p}.
\end{equation}
This model provides a unification of the Starobinsky inflation and Higgs inflation. One can go back to the Higgs model by imposing the constraint
\begin{equation}\label{Hstratum}
e^{\sqrt{\frac{2}{3}}\phi/M_p}=1+\frac{1}{M_p^2}\xi\chi^2
\end{equation}
and the Starobinsky model by
\begin{equation}\label{Sbar}
\chi=0.
\end{equation}

\begin{figure}
  \centering
  \includegraphics[width=0.3\textwidth]{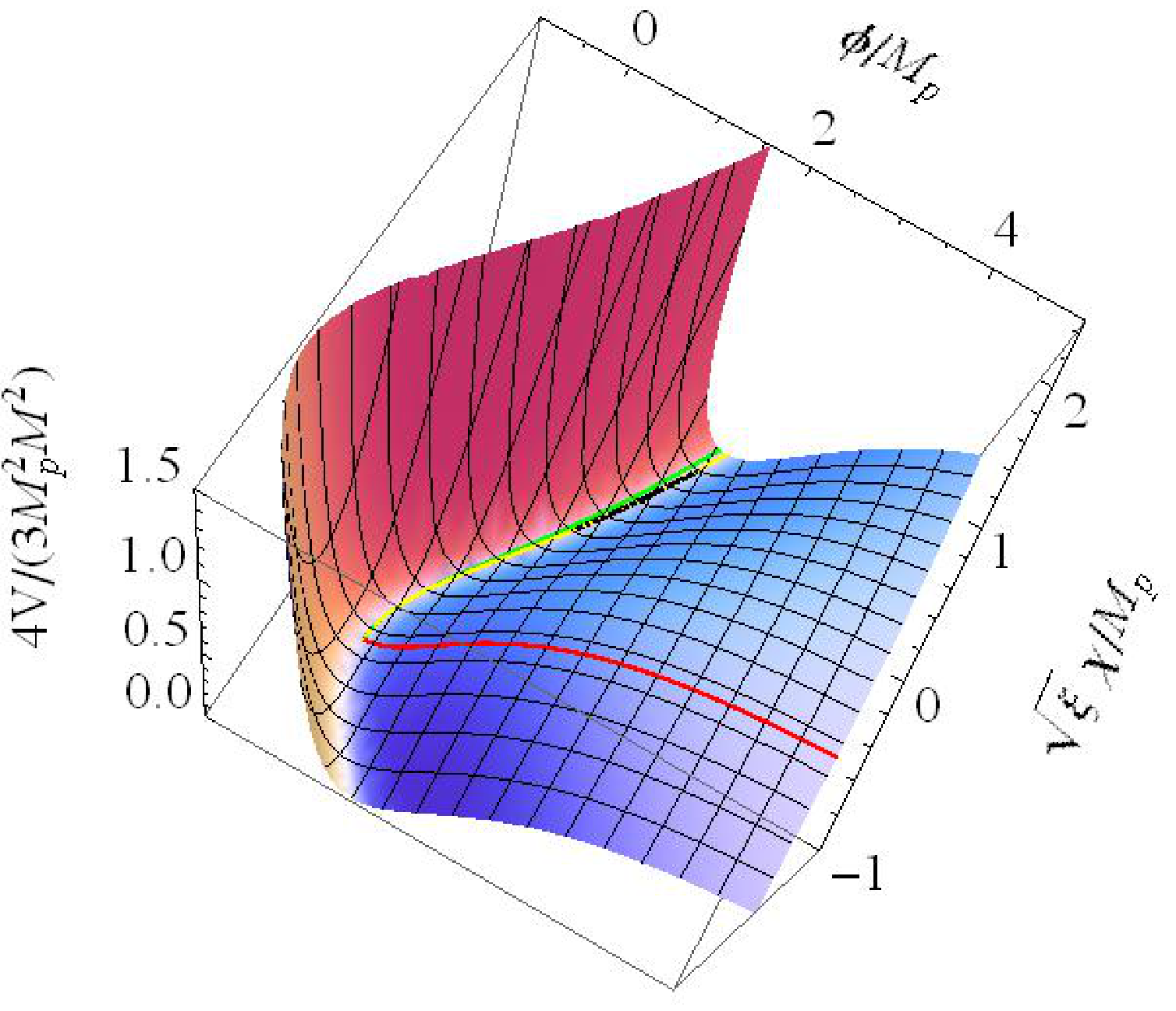}\\
  \includegraphics[width=0.3\textwidth]{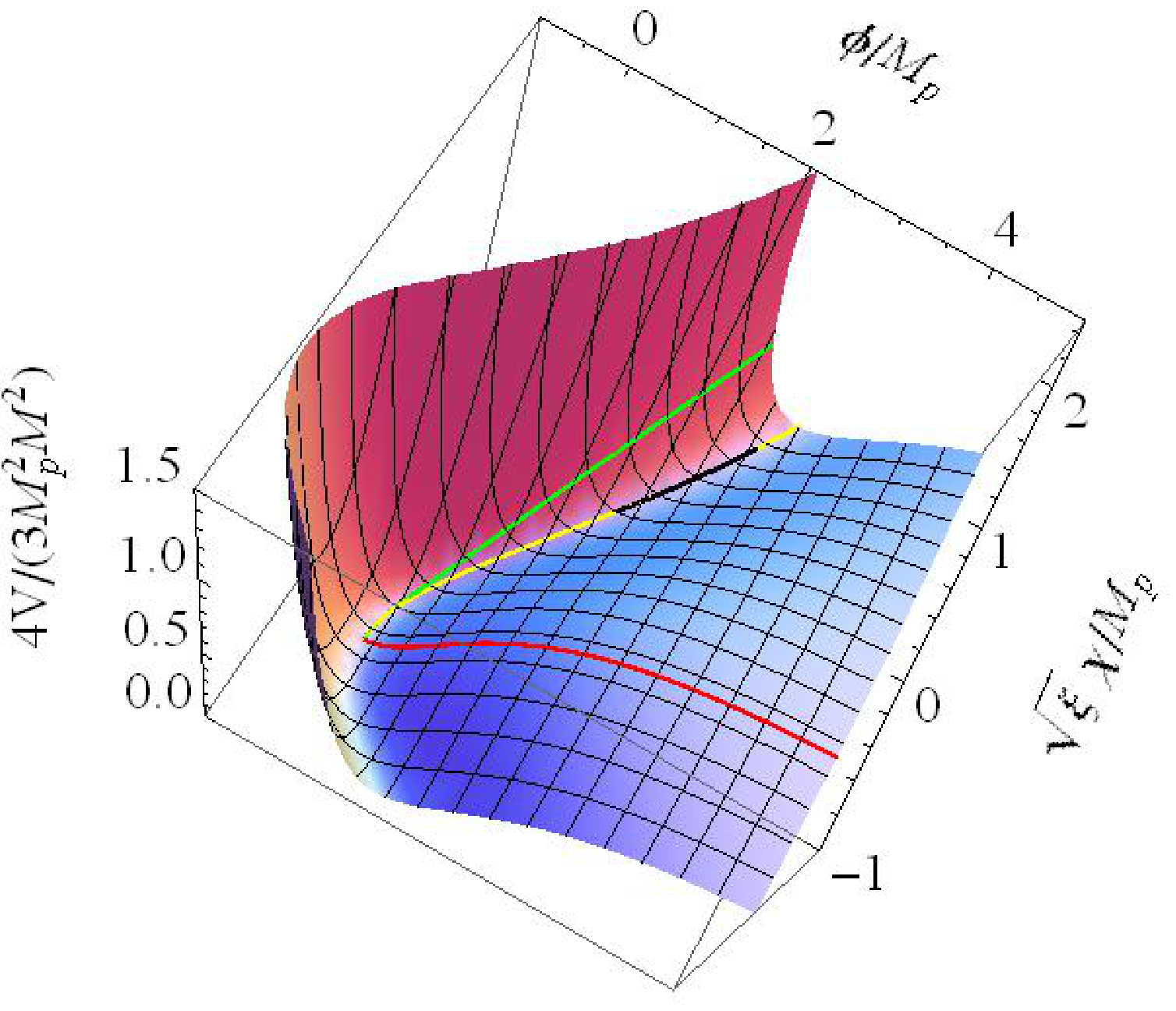}\\
  \includegraphics[width=0.3\textwidth]{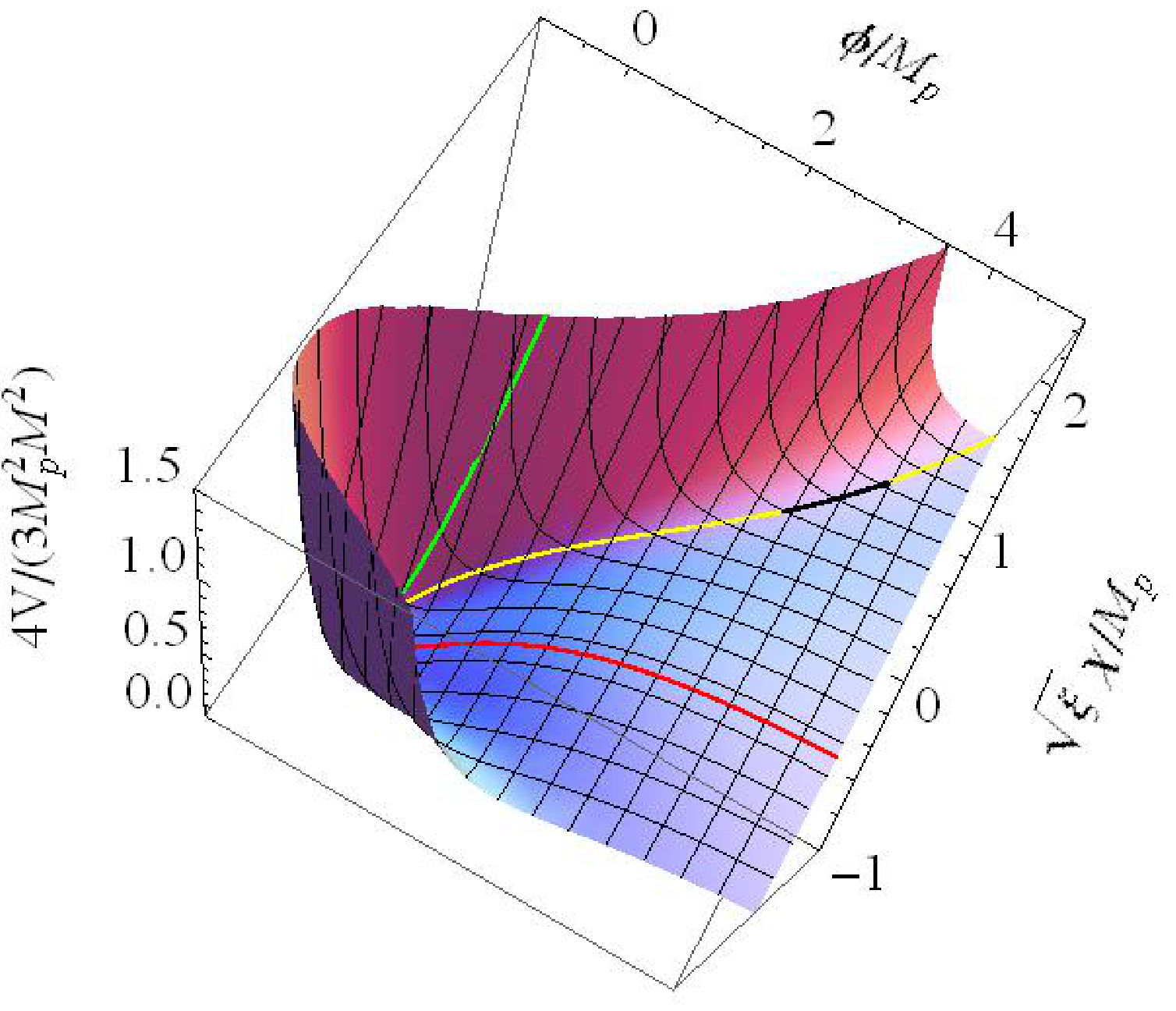}\\
  \caption{(color online). The 3D diagram of potential \eqref{V}. We set concretely $M_p^2\lambda/(3M^{2}\xi^{2})=1/9,1,9$ from top to bottom, but the existence of valley-like landscape is robust to parameter variation. With $M_p^2\lambda/(3M^{2}\xi^{2})$ fixed, varying $M$ and $\xi$ is equivalent to rescaling the $V$-direction and $\chi$-direction respectively. Green, red, yellow trajectories are drawn respectively from Eqs. \eqref{Hstratum}, \eqref{Sbar}, \eqref{valley}. The black thick line in the valley is the field trajectory simulated in Sec. \ref{sect-2f} from equations of motion, starting from $\chi=\chi_{\ast}$ in Eq. \eqref{chistar}, and terminating at the end of inflation $\sqrt{\xi}\chi/M_p=\sqrt{2}$.}\label{fig-V}
\end{figure}

To get a bird view, we present a 3D diagram of potential \eqref{V} in Fig. \ref{fig-V}. It mimics a valley landscape. On the steep side of the valley, the trajectory of Higgs inflation \eqref{Hstratum} is highlighted by a green line like a thin stratum in the cliff. On the other side, the trajectory of Starobinsky inflation \eqref{Sbar} is marked with a red curve along the convex bank akin to a point bar. In this figure, the yellow line represents the pseudo flat direction or the valley of the potential. See the next section for its detailed definition. According to these features in the landscape, we can name the green trajectory \eqref{Hstratum} as Higgs stratum, and the red trajectory \eqref{Sbar} as Starobinsky bar. Apparently, in the full landscape neither the Higgs stratum nor the Srarobinsky bar is an attractive track to run inflation. Instead, the valley itself can serve as a natural route for inflation.

The potential function \eqref{V} depends on three independent parameters $M$, $\xi$, $\lambda$. In Fig. \ref{fig-V}, we have varied only the value of $M_p^2\lambda/(3M^{2}\xi^{2})$, but locked the parameters $M$, $\xi$ into the normalization of $V$ and $\chi$ respectively. This can be understood by reforming Eq. \eqref{V} into
\begin{equation}
\frac{4}{3M_p^2M^2}V(\phi,\chi)=e^{-2\sqrt{\frac{2}{3}}\phi/M_p}\left\{\left[e^{\sqrt{\frac{2}{3}}\phi/M_p}-1-\left(\frac{\sqrt{\xi}\chi}{M_p}\right)^2\right]^2+\frac{M_p^2}{3M^2}\frac{\lambda}{\xi^2}\left(\frac{\sqrt{\xi}\chi}{M_p}\right)^4\right\}.
\end{equation}
Clearly, the parameter $M$ controls the scale of potential diagrams in the $V$-direction, and $\xi$ plays a similar role in the $\chi$-direction. The effect of $M_p^2\lambda/(3M^{2}\xi^{2})$ on the diagram is nontrivial. Displayed in Fig. \ref{fig-V}, as the value of $M_p^2\lambda/(3M^{2}\xi^{2})$ decreases, the valley gets deeper transversely and closer to the Higgs stratum, but in the longitudinal direction it becomes flatter. If $M_p^2\lambda/(3M^{2}\xi^{2})\ll1$, the cliff will be straight, and the yellow line will overlap with the green line, restoring Higgs inflation. When $M_p^2\lambda/(3M^{2}\xi^{2})\gg1$, the valley will form a deep V around the Starobinsky bar, and the yellow line will approach the red curve, recovering Starobinsky inflation.\footnote{Fig. \ref{fig-V} here resembles Fig. 2 of Ref. \cite{Gorbunov:2013dqa} intriguingly. The only difference is that the potential Eq. (22) in Ref. \cite{Gorbunov:2013dqa}, from which that figure was drawn, is an even function of both $\phi$ and $F$. Probably the difference can be attributed to a dilaton involved in Ref. \cite{Gorbunov:2013dqa}.}

In the following, we will locate the valley and study inflationary dynamics along it. Since the potential possesses a reflection symmetry under $\chi\leftrightarrow-\chi$, we will focus on the case $\chi>0$ without loss of generality.

\section{Inflation along the valley}\label{sect-1f}
Model \eqref{act} is a good example of non-canonical two-field inflation, which can be handled numerically or semi-analytically along the way of Refs. \cite{Lalak:2007vi,vandeBruck:2014ata,Wang:2016ipp}. However, we will take a shortcut in the current and next sections, confining the background fields and their perturbations to the valley. The two-field treatment of background and the effect of isocurvature mode will be postponed to Sec. \ref{sect-2f}. We should warn that, when $M_p^2\lambda/(3M^{2}\xi^{2})$ is large, one cannot seriously trust the single-field approximation in this and next sections. The quasi-single field inflation \cite{Chen:2009we,Chen:2009zp} would be a better scenario to study this case. Anyhow, we retain here the example $M_p^2\lambda/(3M^{2}\xi^{2})=9$ of this kind for reference in the future.


In appendix \ref{app-vall}, we propose a method to locate the valley of potential in curved two-field spaces. Applied to model \eqref{act}, it yields the equation of valley
\begin{equation}\label{valley}
\left(V_{\phi}^{2}-e^{\sqrt{\frac{2}{3}}\phi/M_p}V_{\chi}^{2}\right)V_{\phi\chi}=V_{\phi}V_{\chi}\left(V_{\phi\phi}-e^{\sqrt{\frac{2}{3}}\phi/M_p}V_{\chi\chi}\right)+\frac{1}{\sqrt{6}M_p}e^{\sqrt{\frac{2}{3}}\phi/M_p}V_{\chi}^{3}.
\end{equation}
When potential $V(\phi,\chi)$ takes the form \eqref{V}, thanks to the cancellation of higher order terms, this equation turns out a cubic equation with respect to $e^{-\sqrt{\frac{2}{3}}\phi/M_p}$. The coefficients in this cubic equation are lengthy polynomials of $\chi$. The cubic equation has three roots, corresponding to three branches or candidates of valley. Plotting them in the potential diagram, we find only one root coincides with the valley. This is not surprising, because the valley equation \eqref{vacon} is a \emph{necessary} condition for valley. We pick this root as the track for inflation, and depict it with a yellow curve in Fig. \ref{fig-V}. Although the expression of this ``yellow'' root is very complicated, we can still manipulate it numerically, or expand it in series analytically when $\xi\chi^2\ll M_p^2$ or $\xi\chi^2\gg M_p^2$. In this section, we will make use of the series expansion to derive some analytical results, whose error will be assessed in Sec. \ref{sect-data} by higher-order terms numerically. Fortunately, after series expansion in terms of $\xi\chi^2/M_p^2$ or $M_p^2/(\xi\chi^2)$, the coefficients of this yellow root have rational expressions, while the other two roots have irrational coefficients.

If inflatons roll in the valley with a negligible isocurvature perturbation, we can substitute the yellow root of Eq. \eqref{valley} into action \eqref{act} to eliminate the dependence on $\phi$. In the situation $\xi\chi^2\ll M_p^2$, we obtain\footnote{Keep in mind that we have assumed $v^2\ll\chi^2$ in Eq. \eqref{act}, which demands $\xi v^2\ll M_p^2$ in the present case.}
\begin{equation}\label{act1}
\mathcal{S}\approx\int d^4x\sqrt{-\tilde{g}}\left\{\frac{M_p^2}{2}\tilde{R}-\frac{1}{2}\left[1+\frac{1}{M_p^2}\xi(6\xi-1)\chi^2+\cdots\right]\tilde{g}^{\mu\nu}\partial_{\mu}\chi\partial_{\nu}\chi-\frac{1}{4}\lambda\chi^4\left(1-\frac{2}{M_p^2}\xi\chi^2+\cdots\right)\right\},
\end{equation}
in which we have kept the leading-order and next-to-leading-order terms, while ellipses denote higher order terms in $\xi\chi^2\ll M_p^2$. The leading order term dominates if $\xi\lesssim1$. Then the model reduces to a canonical scalar field with the quartic potential, which at the $\chi^2\gg M_p^2$ regime can support slow-roll inflation.\footnote{We are grateful to the kind referee for pointing it out.} Indeed such a regime exists if $\xi\ll1$. As is well known, this model cannot reproduce the observed primordial perturbations, so we will not discuss it any more.

The rest of this paper will be confined to the other situation $\xi\chi^2\gg M_p^2$. In this situation, after substituting the yellow root of Eq. \eqref{valley}, we can rewrite the action \eqref{act} in the kinetic formulation
\cite{Galante:2014ifa} as
\begin{eqnarray}\label{kinact}
\nonumber\mathcal{S}&=&\int d^4x\sqrt{-\tilde{g}}\left[\frac{M_p^2}{2}\tilde{R}-\frac{1}{2}K(\rho)\tilde{g}^{\mu\nu}\partial_{\mu}\rho\partial_{\nu}\rho-V(\rho)\right]\\
&=&\int d^4x\sqrt{-\tilde{g}}\left[\frac{M_p^2}{2}\tilde{R}-\frac{M_p^2}{2}\left(\frac{a_2}{\rho^2}-\frac{a_1}{\rho}+a_0-b_1\rho+\cdots\right)\tilde{g}^{\mu\nu}\partial_{\mu}\rho\partial_{\nu}\rho-V_0(1-c_1\rho+c_2\rho^2-c_3\rho^3+c_4\rho^4+\cdots)\right],
\end{eqnarray}
where $\rho=M_p^2/(\xi\chi^2)$, and ellipses represent higher order terms in $\rho$. Here the leading-order (LO) terms are dictated by
\begin{equation}
V_0=\frac{M_p^4}{4}\frac{\lambda\Lambda}{\xi^2},~~~~a_2=\frac{\Lambda+6\xi}{4\xi},~~~~c_1=2\Lambda.
\end{equation}
For convenience, we have used the notation
\begin{equation}\label{Lambda}
\Lambda=\left(1+\frac{M_p^2}{3M^2}\frac{\lambda}{\xi^2}\right)^{-1}
\end{equation}
which is not greater than one
\begin{equation}\label{theobd1}
\Lambda\leq1.
\end{equation}
The coefficients of next-to-leading-order (NLO) terms can be expressed by $\Lambda$ and $\xi$ as
\begin{eqnarray}
\nonumber a_1&=&\frac{\Lambda\left[\Lambda(2\Lambda-1)+6(5\Lambda-2)\xi+72\xi^2\right]}{4\xi(\Lambda+6\xi)},\\
c_2&=&\frac{\Lambda^2\left[\Lambda(4\Lambda-1)+12(4\Lambda-1)\xi+108\xi^2\right]}{(\Lambda+6\xi)^2}.
\end{eqnarray}
The next-to-next-to-leading-order (NNLO) coefficients are
\begin{eqnarray}
\nonumber a_0&=&\frac{\Lambda^2}{4\xi(\Lambda+6\xi)^3}\left[\Lambda^3(4\Lambda-3)+3\Lambda(41\Lambda^2-28\Lambda-1)\xi+36(31\Lambda^2-20\Lambda+1)\xi^2+432(8\Lambda-3)\xi^3+3888\xi^4\right],\\
c_3&=&\frac{4\Lambda^3}{(\Lambda+6\xi)^4}\left[\Lambda^3(2\Lambda-1)+24\Lambda^2(2\Lambda-1)\xi+9(43\Lambda^2-20\Lambda+1)\xi^2+108(11\Lambda-3)\xi^3+1296\xi^4\right].
\end{eqnarray}

The power series in Eq. \eqref{kinact} will be divergent if $\rho>1$. For this reason, let us restrict our discussion to the region
\begin{equation}\label{theobd2}
\rho<\frac{1}{2}
\end{equation}
until the end of inflation. In our analytical results, we will keep the LO, NLO and NNLO terms in the small-$\rho$ expansion. The next-to-next-to-next-to-leading-order (NNNLO) terms are cumbersome, so they will be omitted in analytical formulae and most numerical results unless otherwise specified. To make the NNLO approximation reliable, we will assume that the NNNLO corrections are smaller than the ordinary slow-roll corrections. In other words, we suppress the NNNLO relative error to less than one percent level
\begin{equation}\label{theobd3}
\rho^3<0.01.
\end{equation}
It is easy to check that Eq. \eqref{theobd2} is looser than this condition.

For succinctness and explicitness, we introduce the \emph{uncalibrated} e-folding number
\begin{equation}\label{unefolds}
\mathcal{N}=\frac{a_2}{\rho c_1}=\frac{\Lambda+6\xi}{8\xi\Lambda\rho}
\end{equation}
which indicates that the small-$\rho$ expansion is equivalent to a large-$\mathcal{N}$ expansion
\begin{equation}\label{rho}
\rho=\frac{\Lambda+6\xi}{8\mathcal{N}\xi\Lambda}.
\end{equation}
The right hand side becomes roughly $1/(\mathcal{N}\Lambda)$ if $\Lambda/\xi\leq1$, and tends to $1/(8\mathcal{N}\xi)$ when $\Lambda/\xi\gg1$. With the aid of Eq. \eqref{rho}, one can rewrite Eqs. \eqref{theobd2}, \eqref{theobd3} as
\begin{equation}\label{err}
\frac{\Lambda+6\xi}{8\mathcal{N}\xi\Lambda}<\frac{1}{2},~~~~\left(\frac{\Lambda+6\xi}{8\mathcal{N}\xi\Lambda}\right)^3<0.01.
\end{equation}
Together with Eq. \eqref{theobd1}, they set the theoretical consistency conditions to our analytical study in this paper. The notation $\mathcal{N}$ should not be confused with the \emph{true} e-folding number $N=\ln(a_e/a)$. They are related by
\begin{eqnarray}
\nonumber N&\simeq&\int_{\rho_e}^{\rho}\frac{KV}{M_p^2V_{\rho}}d\rho\\
\nonumber&\approx&\mathcal{N}-\frac{a_2}{\rho_e c_1}+\frac{3}{4}\ln\frac{\rho}{\rho_e}+\frac{3\Lambda\left[(5\Lambda^2-2\Lambda-1)+12(3\Lambda-1)\xi+72\xi^2\right]}{8(\Lambda+6\xi)^2}\left(\rho-\rho_e\right)\\
&=&\mathcal{N}-\frac{\Lambda+6\xi}{4\xi\Lambda}+\frac{3}{4}\ln\frac{\Lambda+6\xi}{4\mathcal{N}\xi\Lambda}+\frac{3\Lambda\left[(5\Lambda^2-2\Lambda-1)+12(3\Lambda-1)\xi+72\xi^2\right]}{8(\Lambda+6\xi)^2}\left(\frac{\Lambda+6\xi}{8\mathcal{N}\xi\Lambda}-\frac{1}{2}\right).
\end{eqnarray}
In the last step, we have assumed that the inflation stops at $\rho_e=1/2$ in accordance with Eq. \eqref{theobd2}. The quantities with a subscript $e$ are evaluated at the end of inflation. We always use $\simeq$ to denote equivalence up to terms suppressed by the slow-roll parameter $\epsilon=-\dot{H}/H^2$, and use $\approx$ to denote equivalence up to NNNLO terms in the small-$\rho$ expansion. From this relation, it is clear that the uncalibrated e-folding number $\mathcal{N}$ is the leading-order approximation of the true e-folding number $N$. To understand this directly, one may compare Eq. \eqref{unefolds} in this paper with Eq. (8) in Ref. \cite{Galante:2014ifa} (or Eq. (2.9) in the arXiv version) setting $p=2$.

Following Ref. \cite{Galante:2014ifa}, one can calculate amplitude of curvature power spectrum, the scalar spectral index and the tensor-to-scalar ratio order by order in $\rho$, or equivalently in $1/\mathcal{N}$. After straightforward computation accurate to NNLO terms, we find they are given by
\begin{eqnarray}
\nonumber A_s&\simeq&\frac{KV^3}{12\pi^2M_p^6V_{\rho}^2}\\
\nonumber&\approx&\frac{\mathcal{N}^2\lambda\Lambda}{12\pi^2\xi(\Lambda+6\xi)}-\frac{\mathcal{N}\lambda\Lambda\left[\Lambda+6(\Lambda+2)\xi+72\xi^2\right]}{96\pi^2\xi^2(\Lambda+6\xi)^2}\\
&&+\frac{\lambda\Lambda}{768\pi^2\xi^3(\Lambda+6\xi)^3}\left[\Lambda^2+3\Lambda\left(5\Lambda^2+7\right)\xi+36(7\Lambda^2+3\Lambda+2)\xi^2+432(4\Lambda+1)\xi^3+3888\xi^4\right],\label{As}\\
\nonumber n_s-1&\simeq&-\frac{M_p^2K_{\rho}V_{\rho}}{K^2V}-\frac{3M_p^2V_{\rho}^2}{KV^2}+\frac{2M_p^2V_{\rho\rho}}{KV}\\
\nonumber&\approx&-\frac{2}{\mathcal{N}}-\frac{\Lambda+6(3\Lambda+2)\xi+144\xi^2}{8\mathcal{N}^2\xi(\Lambda+6\xi)}\\
&&+\frac{1}{64\mathcal{N}^3\xi^2(\Lambda+6\xi)^2}\left[\Lambda^2+6\Lambda(10\Lambda^2-5\Lambda+2)\xi+36(21\Lambda^2-9\Lambda-1)\xi^2+216(13\Lambda-8)\xi^3\right],\label{ns}\\
\nonumber r&\simeq&\frac{8M_p^2V_{\rho}^2}{KV^2}\\
\nonumber&\approx&\frac{12}{\mathcal{N}^2}\left(1+\frac{\Lambda}{6\xi}\right)+\frac{\Lambda(1-2\Lambda)+6(2-3\Lambda)\xi}{4\mathcal{N}^3\xi^2}\\
&&+\frac{1}{32\mathcal{N}^4\xi^3(\Lambda+6\xi)}\left[\Lambda^3(4\Lambda-3)+3\Lambda(23\Lambda^2-20\Lambda+1)\xi+36(9\Lambda^2-10\Lambda+2)\xi^2\right].\label{r}
\end{eqnarray}
As mentioned above, in this paper, we will impose the condition \eqref{theobd3} so that the NNNLO corrections are smaller than the ordinary slow-roll corrections. The NNNLO terms are not shown here, but they will be evaluated numerically in Sec. \ref{sect-data} to confirm that their corrections are less than one percent.

In the leading order, Eqs. \eqref{As}, \eqref{ns}, \eqref{r} have the same behavior as $\alpha$-attractors \cite{Ferrara:2013rsa,Kallosh:2013yoa,Linde:2015uga}, \footnote{For a recent development, see Ref. \cite{Karananas:2016kyt} and references therein.}
\begin{equation}\label{alpha}
A_s=\frac{\mathcal{N}^2\lambda\Lambda}{12\pi^2\xi(\Lambda+6\xi)},~~~n_s-1=-\frac{2}{\mathcal{N}},~~~~r=\frac{12\alpha}{\mathcal{N}^2},~~~~\alpha=1+\frac{\Lambda}{6\xi}.
\end{equation}
Eliminating $\xi$ with Eq. \eqref{unefolds}, we can combine them to write down a relation
\begin{equation}\label{rbd}
r=8\Lambda\rho(1-n_s).
\end{equation}
Remember that both $\Lambda$ and $\rho$ are bounded from above, so this relation will impose an upper bound on the tensor mode in the LO. Most generally and most loosely, we have Eq. \eqref{theobd1} and $\rho<1$, then the LO bound is $r<8(1-n_s)$.

In this paper, more conservative Eqs. \eqref{theobd2}, \eqref{theobd3} are assumed for an accurate study. Substituted into Eq. \eqref{rbd}, they give a more conservative bound $r<1.72(1-n_s)$. In the coming section, a large $r$ of the same order will be found in the accurate NNLO simulations. Such an upper bound value is encouraging, for it is close to the precision of ongoing CMB observations.

We mention two special cases in passing. First, in the limit $M_p^2/M^2\rightarrow0$, if $\xi\gg1$, then Eq. \eqref{alpha} tends to the result of Higgs model,
\begin{equation}
A_s=\frac{N^2}{72\pi^2M_p^2}\frac{\lambda}{\xi^2},~~~~n_s=1-\frac{2}{N},~~~~r=\frac{12}{N^2}.
\end{equation}
Second, for the special case $\xi=1$, if $\Lambda\ll1$, it is better to end the inflation at $\rho_e\Lambda=1/2$. This yields the leading-order result
\begin{equation}
A_s=\frac{N^2M^2}{24\pi^2M_p^2},~~~~n_s=1-\frac{2}{N},~~~~r=\frac{12}{N^2}
\end{equation}
the same as the predictions of Starobinsky model. Note in both cases, the difference between uncalibrated e-folding number and true e-folding number is small
\begin{equation}
N\sim\mathcal{N}-\frac{3}{2}+\frac{3}{4}\ln\frac{3}{2\mathcal{N}}\sim\mathcal{N}.
\end{equation}

From the LO Eq. \eqref{alpha}, one can deduce that $A_sr=\lambda\Lambda/(6\pi^2\xi^2)$. By this equality and Eq. \eqref{Planck15} from Planck 2015 results \cite{Ade:2015xua}, we obtain a rough bound
\begin{equation}\label{tune}
\frac{3\xi^2}{\lambda}+\frac{M_p^2}{M^2}=\frac{1}{2\pi^2A_sr}>2.14\times10^{8}.
\end{equation}
As indicated by this bound, fine-tuning is unavoidable in viable models. One has to tune either the Higgs couplings just as usually done in Higgs inflation model, or the coefficient of $R^2$ term like Starobinsky model. Or both.

\section{Comparison with observational data}\label{sect-data}
Eqs. \eqref{ns}, \eqref{r} connect three model parameters $\Lambda$, $\xi$, $\mathcal{N}$ to two cosmological parameters $n_s$, $r$, while Eq. \eqref{As} involves two more parameters $\lambda$ and $A_s$. The model parameters $\Lambda$, $\xi$, $\mathcal{N}$ are limited by theoretical consistency conditions \eqref{theobd1}, \eqref{err}. With regard to cosmological parameters, the $68\%$ C.L. limits on $A_s$, $n_s$ and the $95\%$ C.L. limit on $r_{0.002}$ from Planck 2015 results \cite{Ade:2015xua} are
\begin{equation}\label{Planck15}
10^9A_s=2.139\pm0.063,~~~~n_s=0.9677\pm0.0060,~~~~r_{0.002}<0.114.
\end{equation}
Firstly, these limits can be combined to constrain our model parameters. Secondly, if we set the NNLO values of $A_s$, $n_s$ to their best-fit values, then relations between other parameters can be made visible. What is more, replacing the left hand side of Eqs. \eqref{As}, \eqref{ns} with the best-fit values, we can numerically compare the LO, NLO, NNLO, NNNLO results of $A_s$, $n_s$, $r$. These are what we plan to do in the current section.

\begin{figure}
  \centering
  \includegraphics[width=0.22\textwidth]{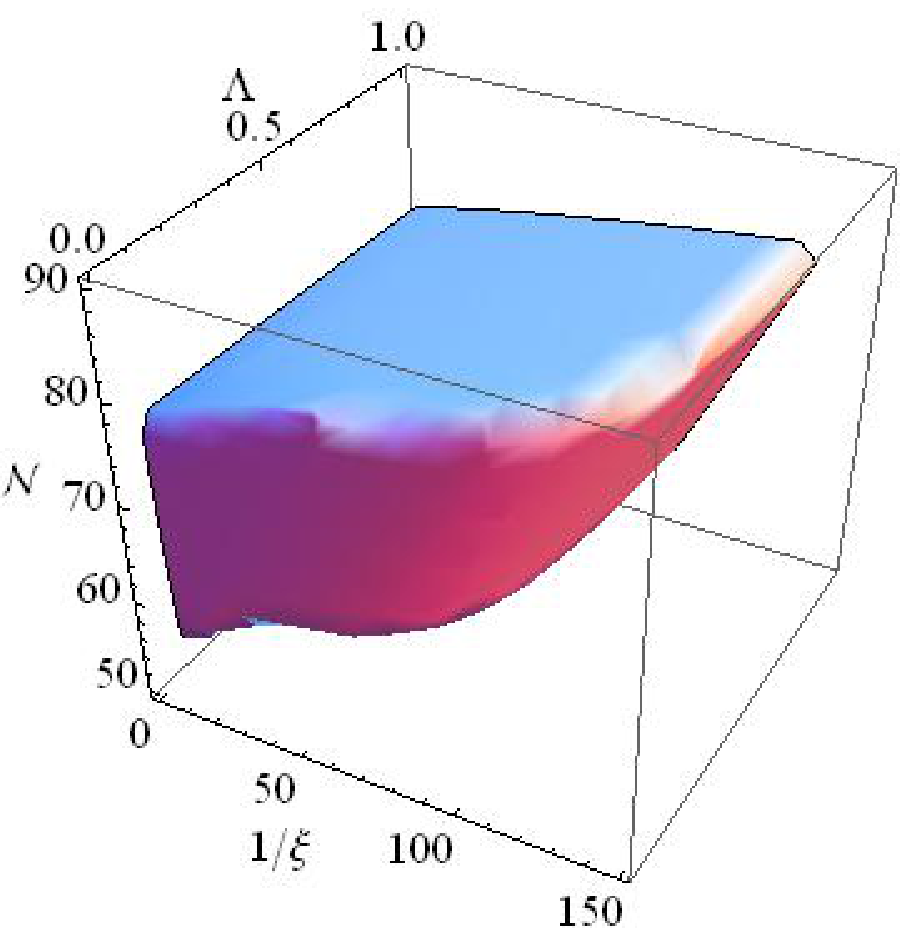}\includegraphics[width=0.22\textwidth]{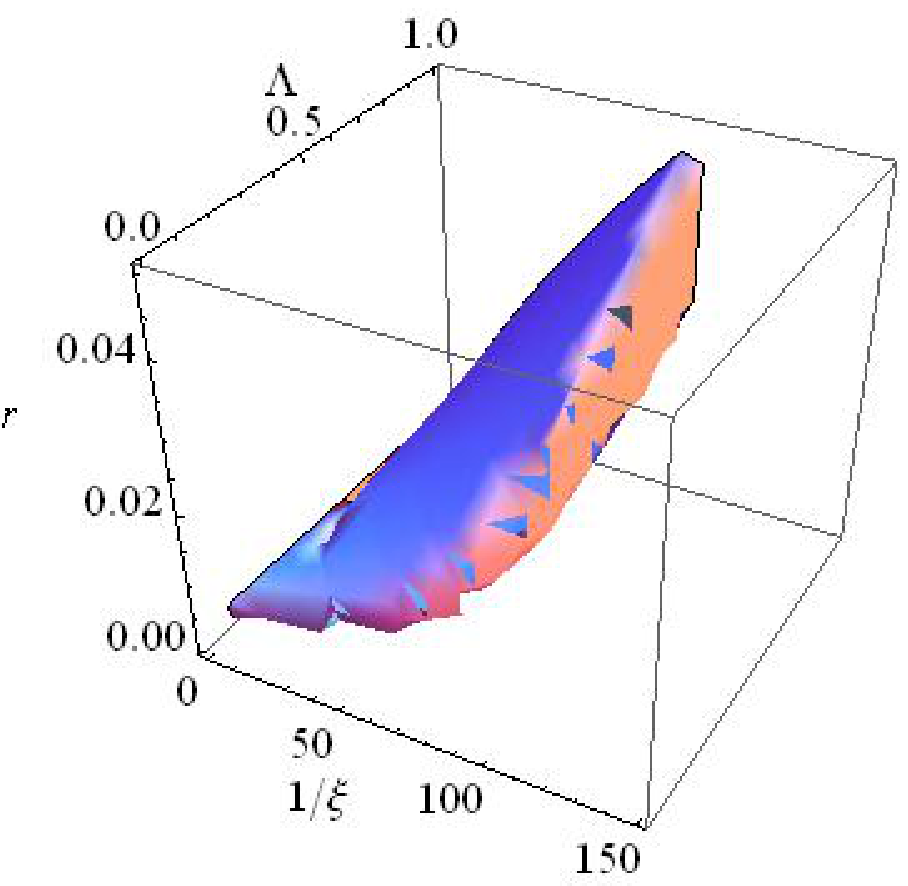}\\
  \caption{(color online). Observational and theoretical constraints on parameters $\Lambda$, $\xi$, $\mathcal{N}$ in the left panel, and on parameters $\Lambda$, $\xi$, $r$ in the right panel. We impose the $68\%$ C.L. limits on $n_s$ and the $95\%$ C.L. limit on $r_{0.002}$ from Planck 2015 results \cite{Ade:2015xua} through the NNLO Eqs. \eqref{ns}, \eqref{r}. The condition $A_s/\lambda>0$ is imposed as well through the NNLO Eq. \eqref{As}. Theoretical constraints are Eqs. \eqref{theobd1}, \eqref{theobd2}, \eqref{theobd3}. The allowed 3D parameter space is enclosed in the colored chunk. The small rips in the right panel are caused by numerical errors in our simulation.}\label{fig-chunk}
\end{figure}

Although the full model-parameter space ($\Lambda$, $\xi$, $\mathcal{N}$, $\lambda$) is 4-dimensional, the last parameter $\lambda$ can be easily disentangled, because it enters in observational and theoretical conditions through Eq. \eqref{As} exclusively. Combining observational constraints \eqref{Planck15} with theoretical constraints \eqref{theobd1}, \eqref{err}, we derived the allowed space of model parameters $\Lambda$, $\xi$, $\mathcal{N}$. It is depicted by the colored chunk in the left panel of Fig. \ref{fig-chunk}. Subject to the same constraints, the allowed space of parameters $\Lambda$, $\xi$, $r$ is depicted by the colored chunk in the right panel of Fig. \ref{fig-chunk}. The NNLO upper bound on $r$, which can be seen from the right panel, is not far from our LO estimation Eq. \eqref{rbd}.

\begin{figure}
  \centering
  \includegraphics[width=0.22\textwidth]{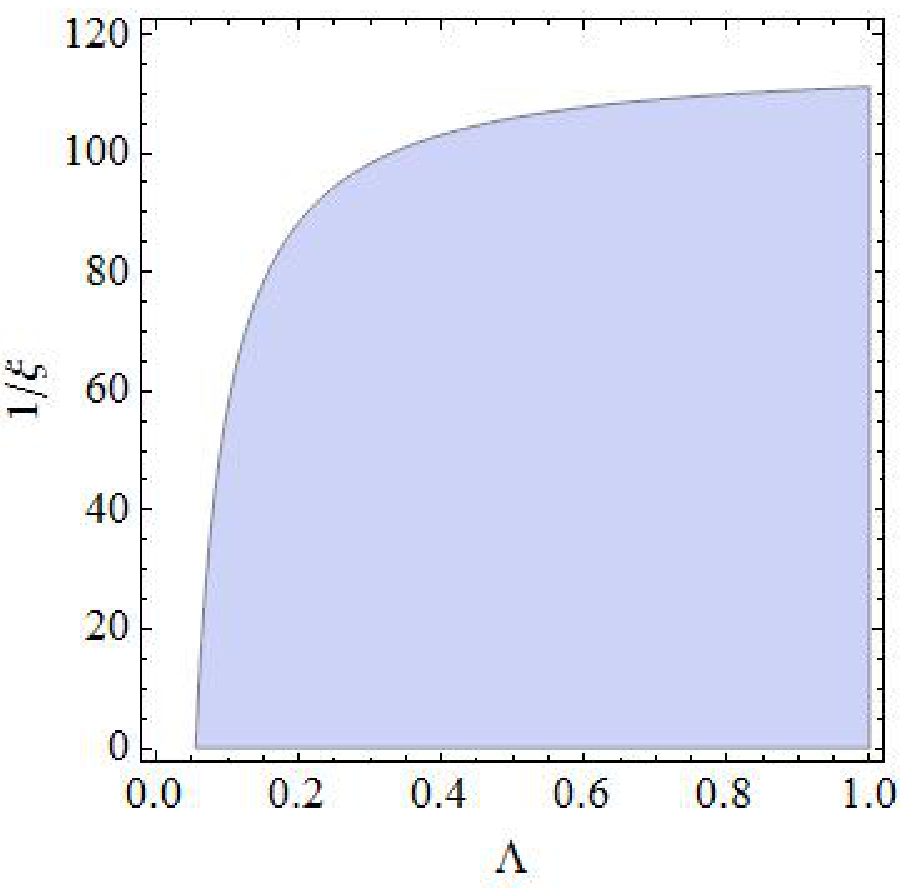}\includegraphics[width=0.22\textwidth]{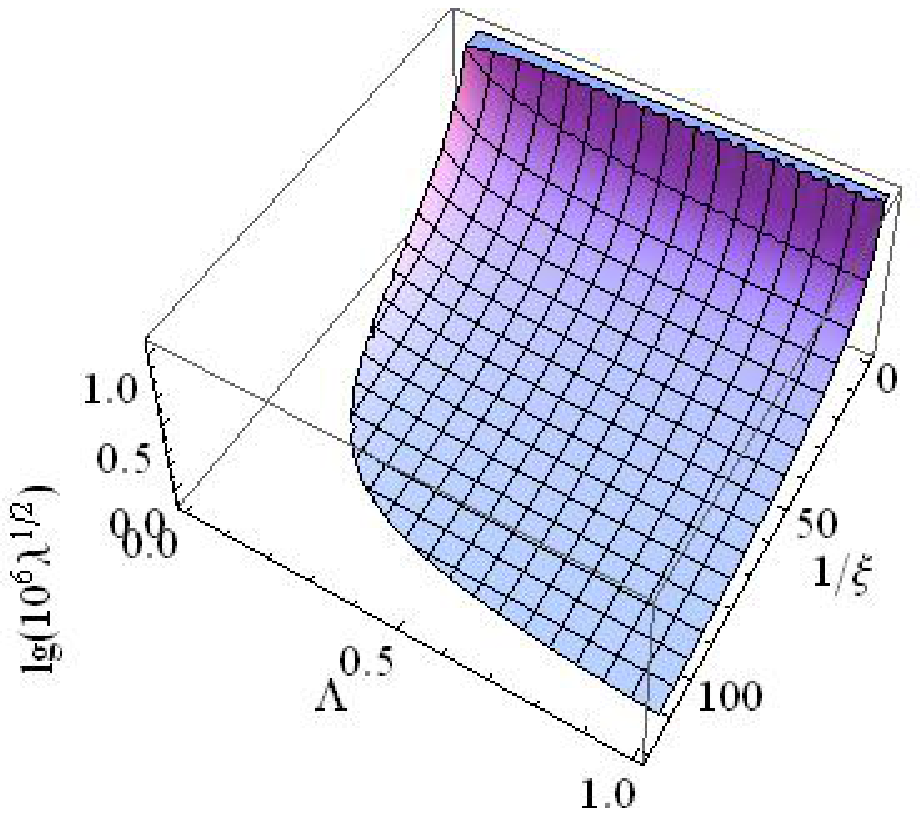}\\
  \includegraphics[width=0.22\textwidth]{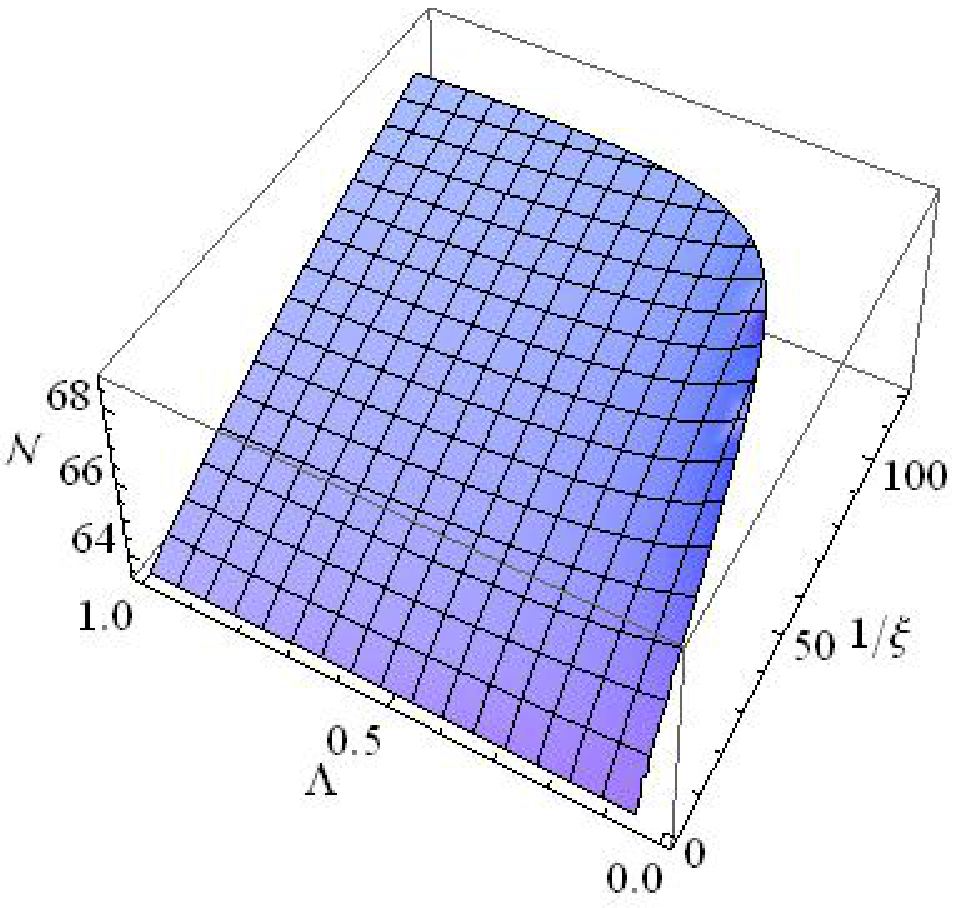}\includegraphics[width=0.22\textwidth]{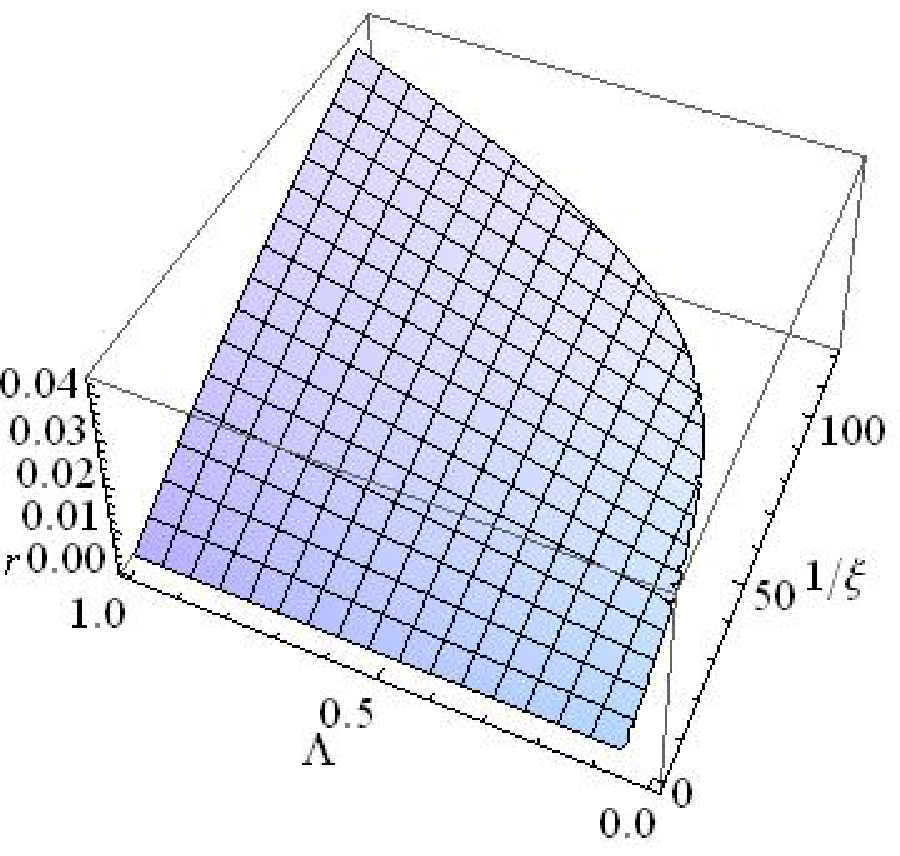}\\
   \caption{(color online). Observational and theoretical constraints on parameters $\Lambda$, $\xi$, $\lambda$, $\mathcal{N}$, $r$. We impose the best-fit values of $A_s$, $n_s$ and the $95\%$ C.L. limit on $r_{0.002}$ from Planck 2015 results \cite{Ade:2015xua} through the NNLO Eqs. \eqref{As}, \eqref{ns}, \eqref{r}. Theoretical constraints are Eqs. \eqref{theobd1}, \eqref{theobd2}, \eqref{theobd3}. The allowed region of ($\Lambda$, $\xi$) is painted blue in the upper-left panel. The surfaces in upper-right, lower-left, lower-right panels depict the dependence of $\lambda$, $\mathcal{N}$, $r$ respectively on $\Lambda$ and $\xi$. }\label{fig-bf}
\end{figure}

With $A_s$, $n_s$ fixed to the best-fit values, the relations between other parameters will be simplified. Under this assumption, the allowed region of ($\Lambda$, $\xi$) is painted in the upper-left panel of Fig. \ref{fig-bf}. In the allowed region, the other panels of Fig. \ref{fig-bf} exhibit how parameters $\lambda$, $\mathcal{N}$, $r$ depend on $\Lambda$ and $\xi$ according to Eqs. \eqref{As}, \eqref{ns}, \eqref{r}. The lower panels are nothing else but the cross-section of the chunks in Fig. \ref{fig-chunk} at $n_s=0.9677$.

When drawing Figs. \ref{fig-chunk} and \ref{fig-bf}, we have taken the theoretical restrictions \eqref{theobd1}, \eqref{err} into account, while Eq. \eqref{err} is equivalent to Eqs. \eqref{theobd2}, \eqref{theobd3}. As an aside, we find the restriction on ($\Lambda$, $\xi$) comes mainly from Eq. \eqref{theobd3}. We have tried to replace the right hand side of Eq. \eqref{theobd3} with $1/8$, and found the lower bound on $\Lambda$ is relaxed to about $0.024$, while the upper bound for $1/\xi$ goes to around $275$.

\begin{figure}
  \centering
  \includegraphics[width=0.3\textwidth]{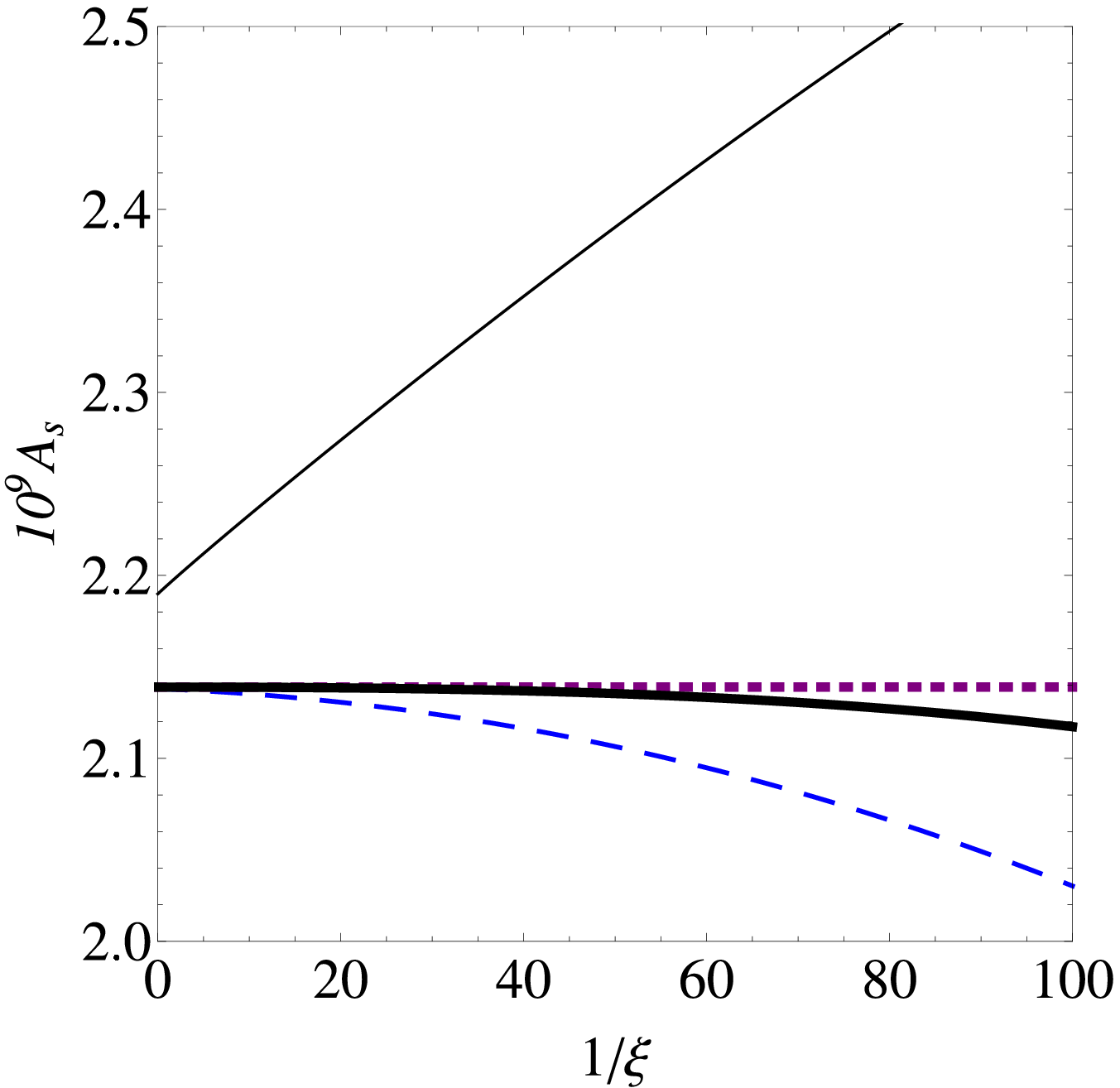}\includegraphics[width=0.3\textwidth]{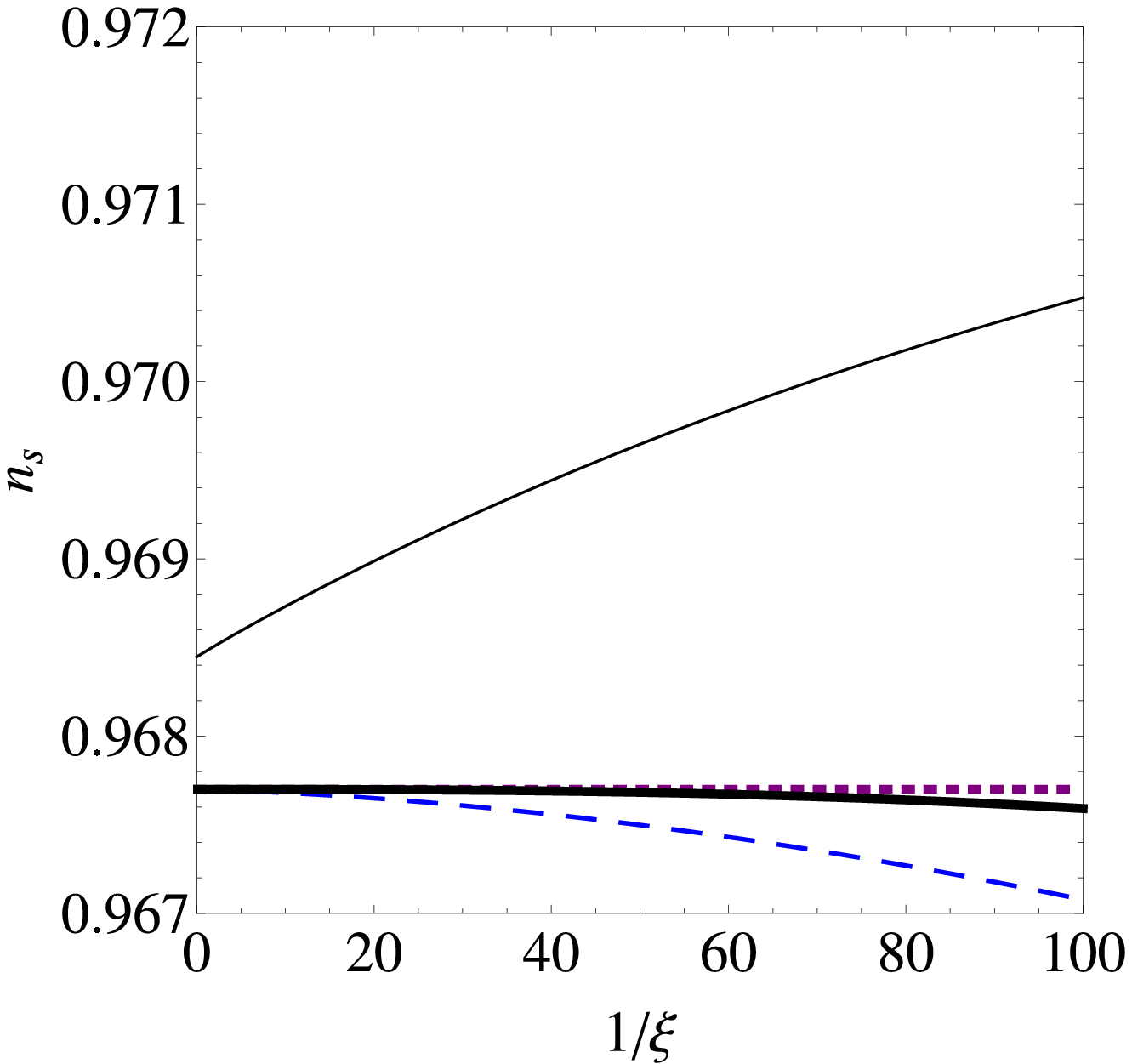}\includegraphics[width=0.3\textwidth]{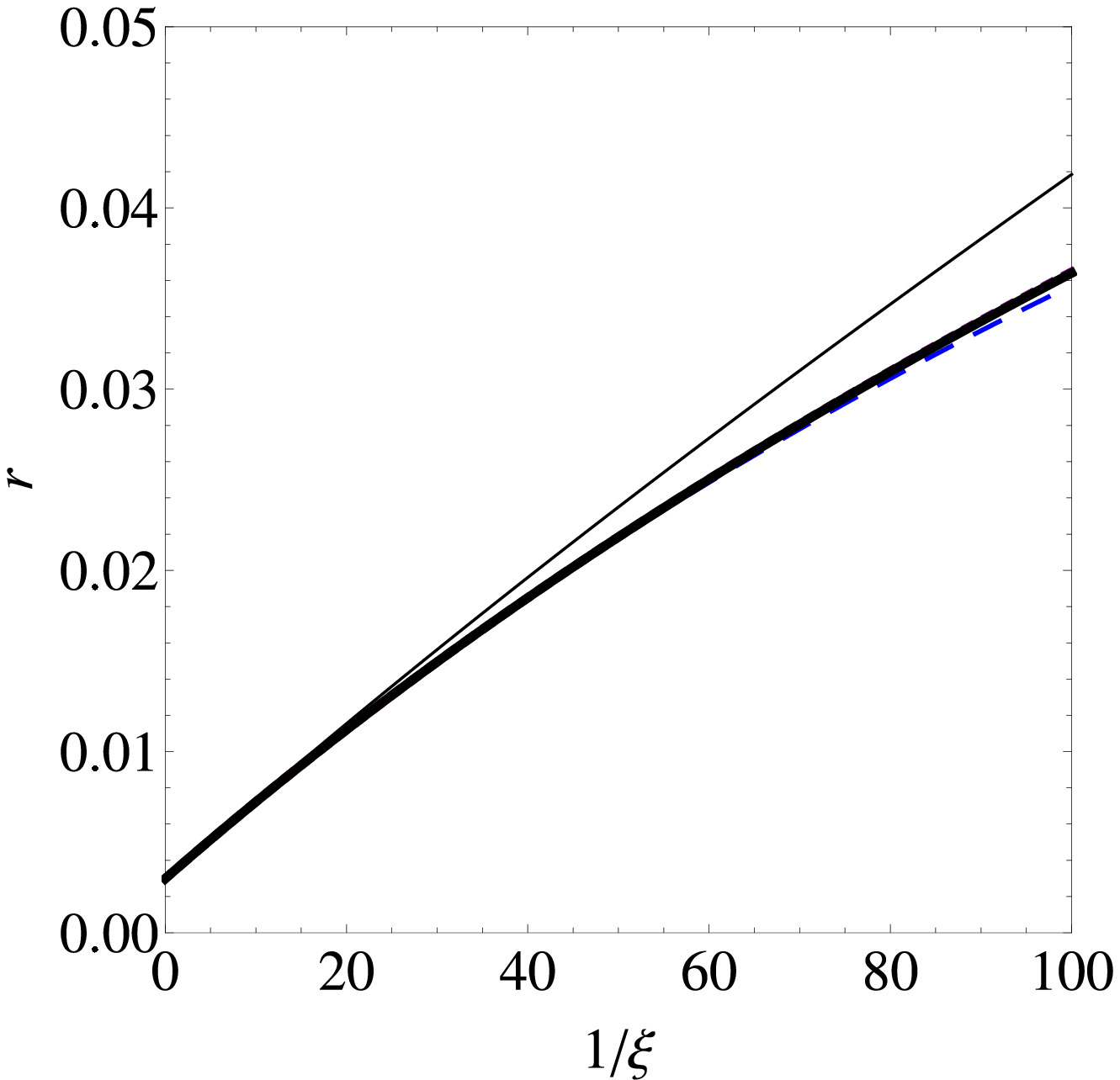}\\
  \includegraphics[width=0.3\textwidth]{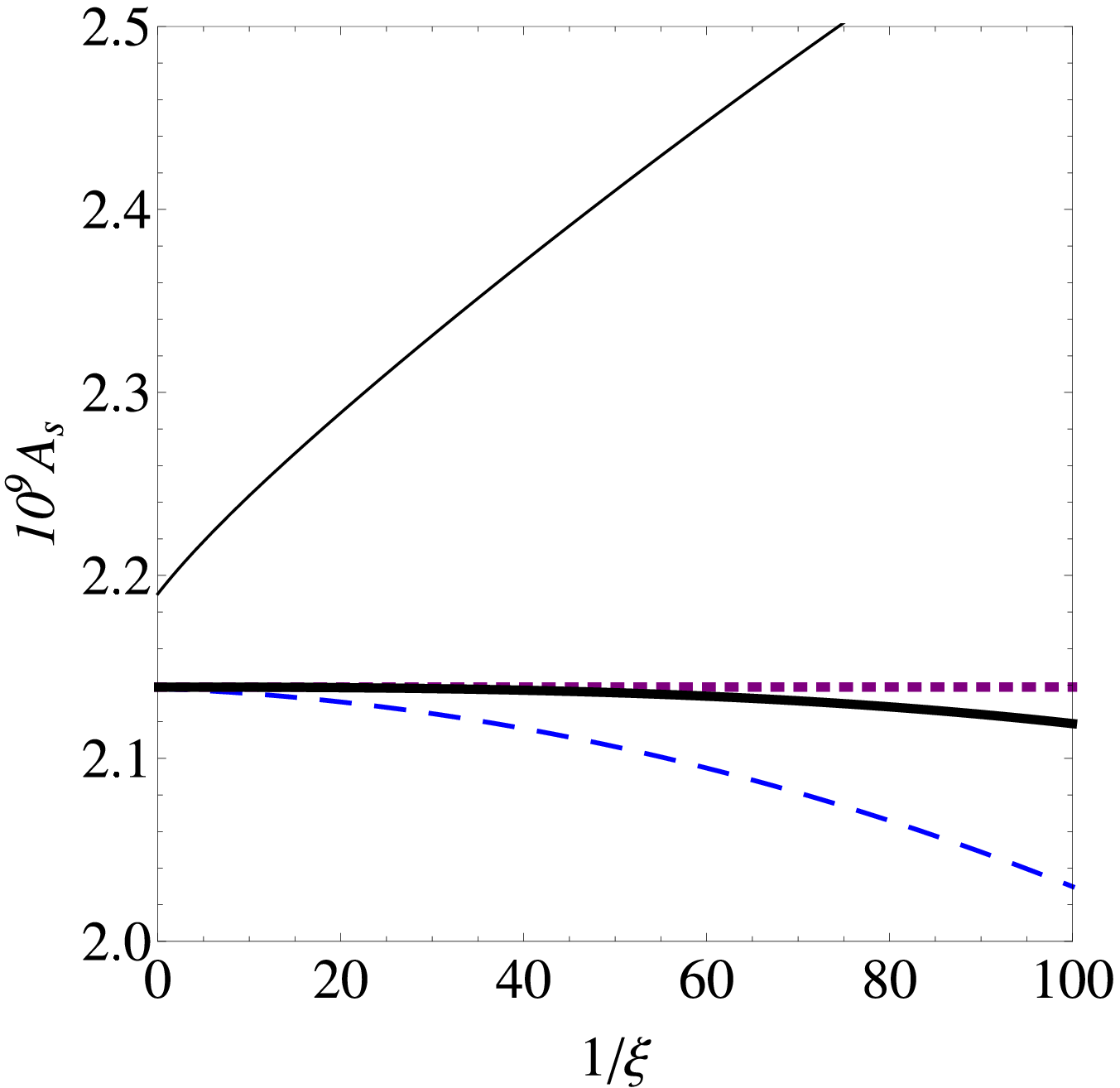}\includegraphics[width=0.3\textwidth]{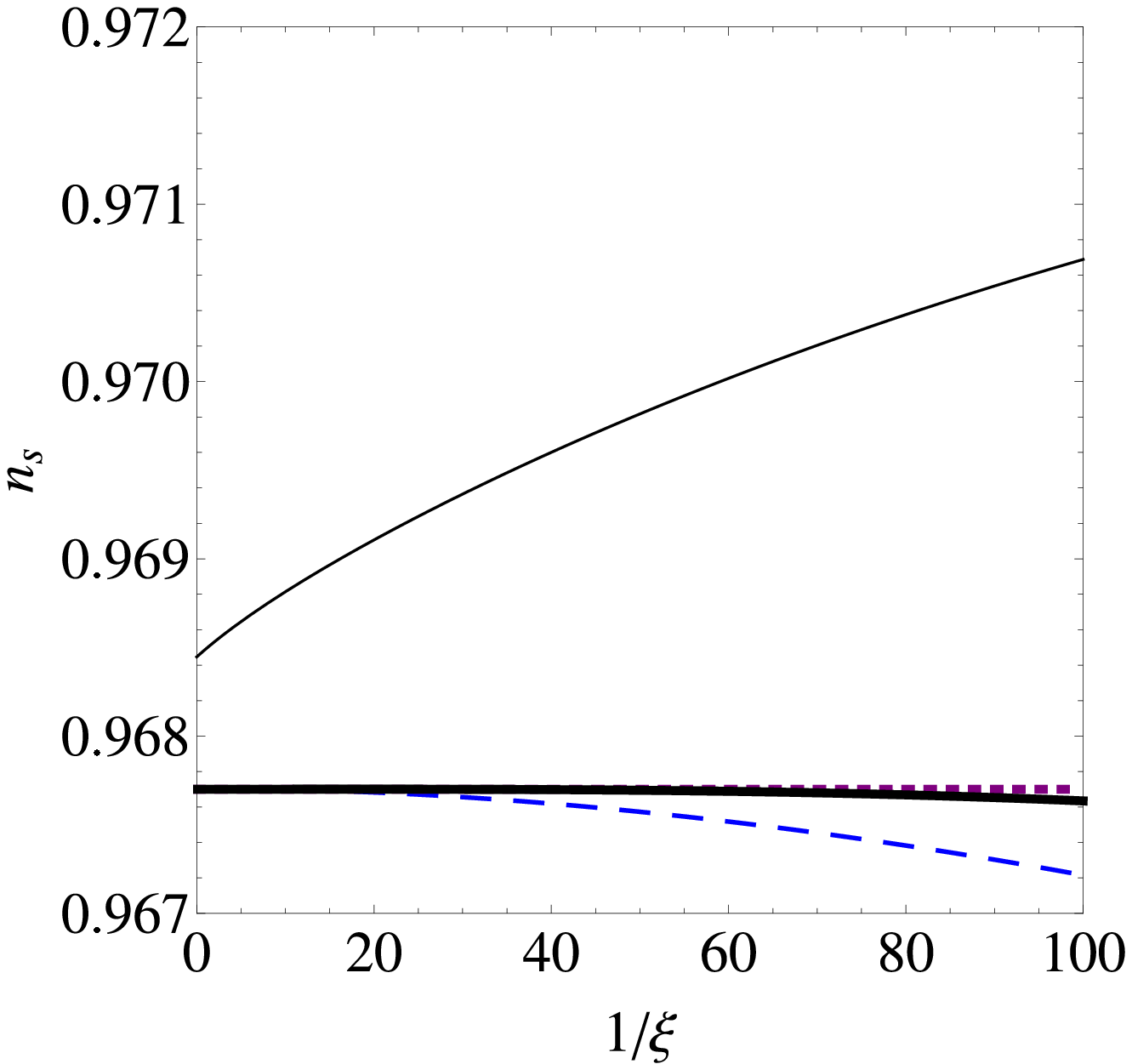}\includegraphics[width=0.3\textwidth]{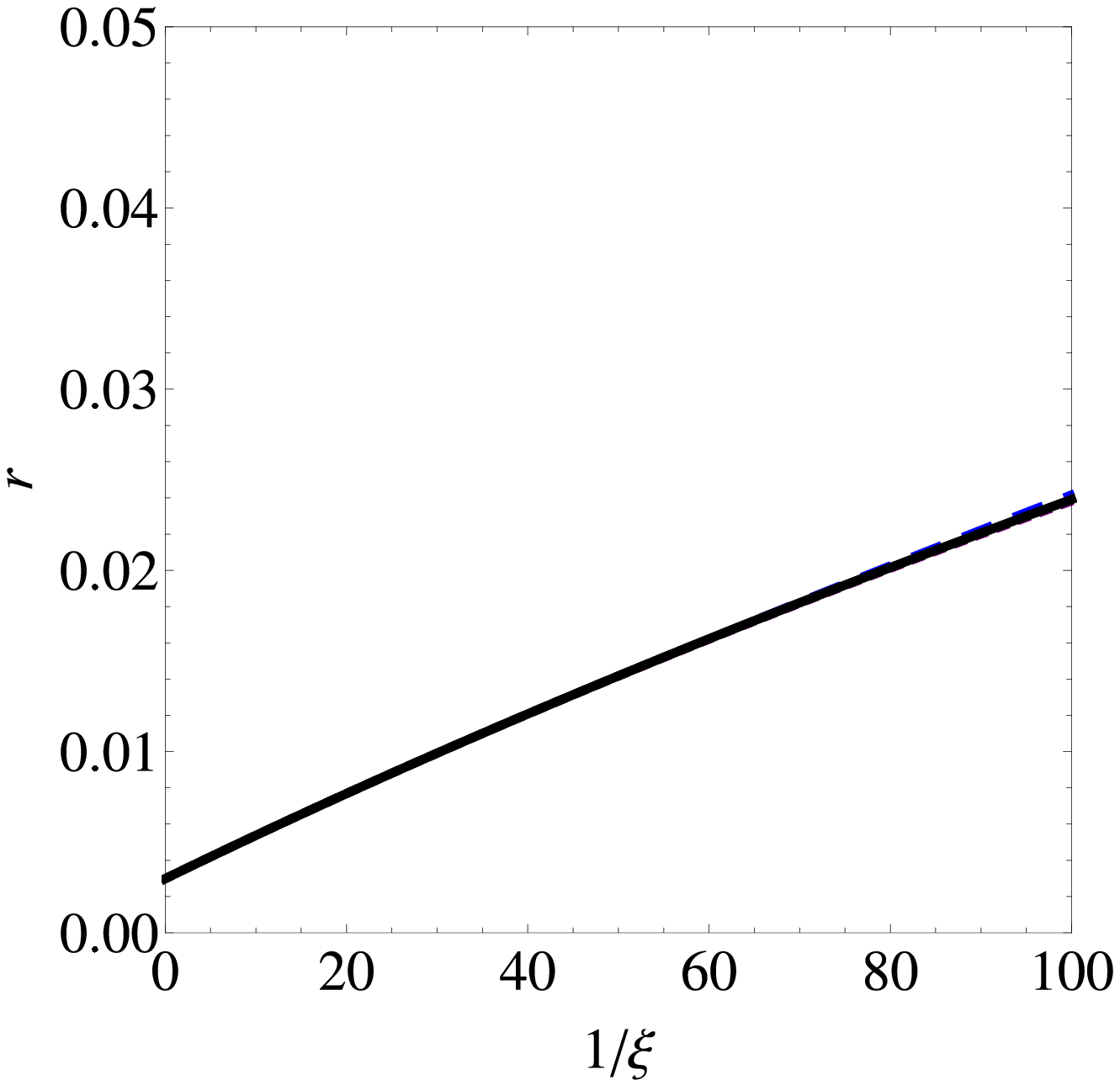}\\
  \includegraphics[width=0.3\textwidth]{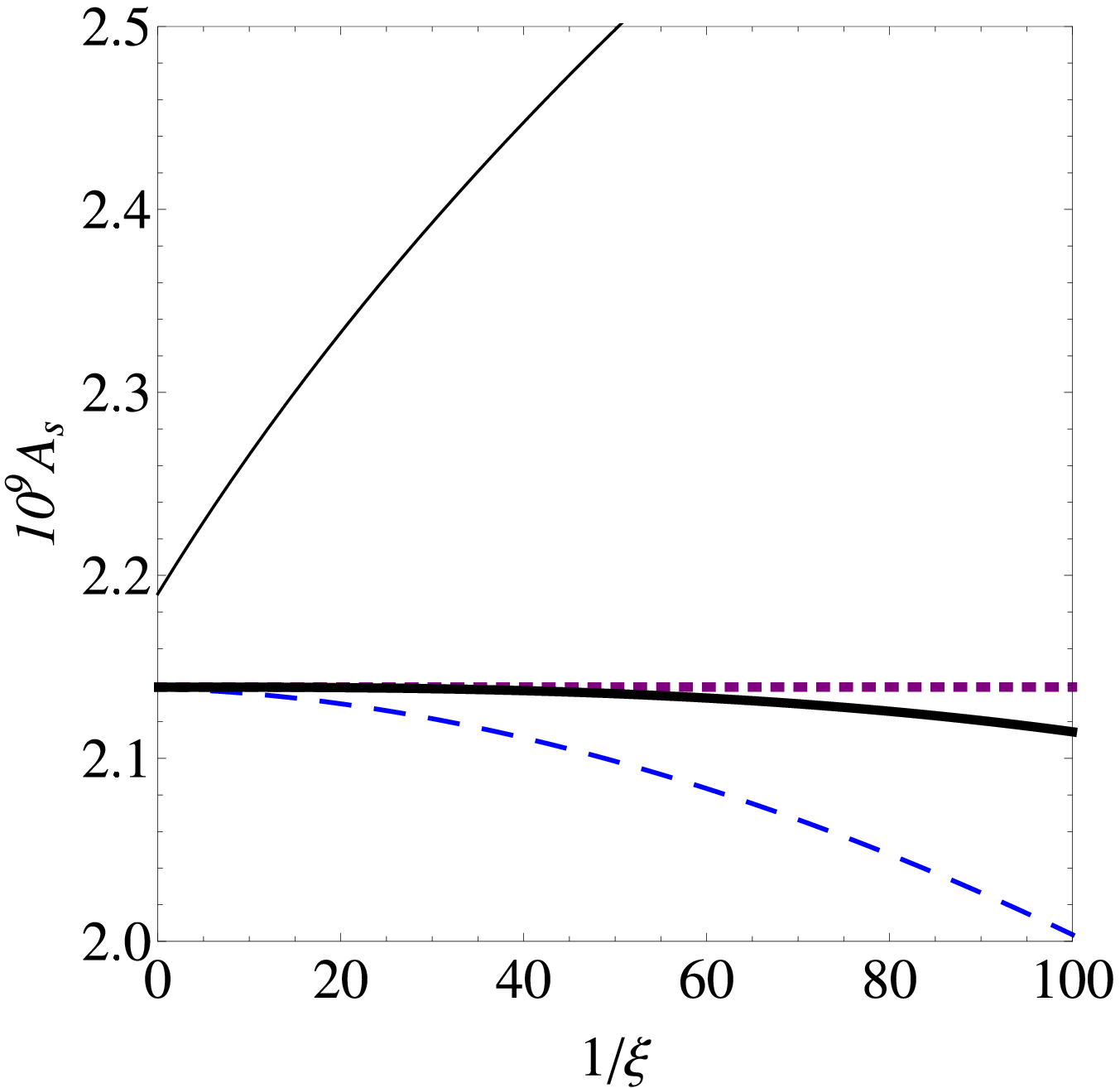}\includegraphics[width=0.3\textwidth]{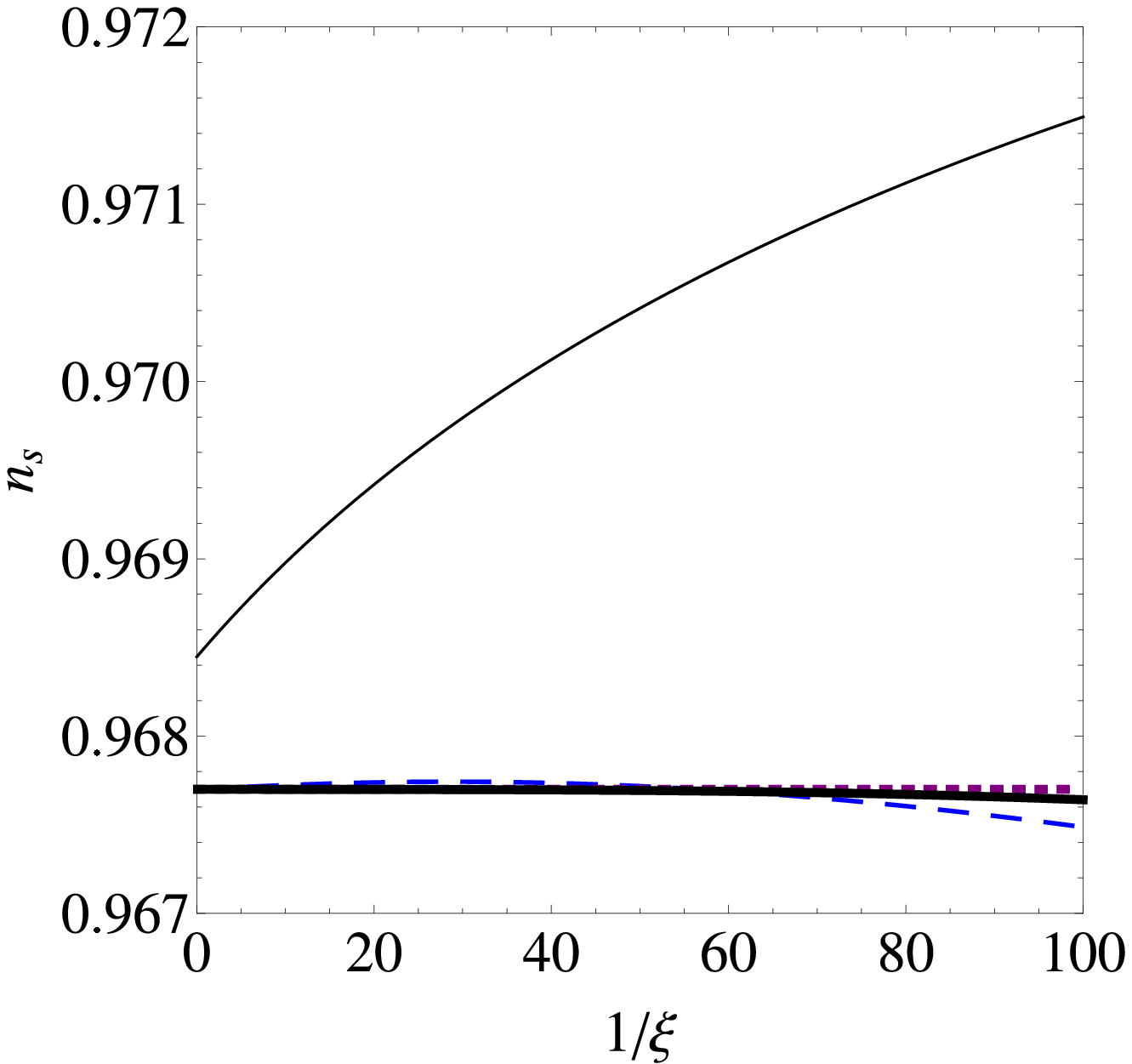}\includegraphics[width=0.3\textwidth]{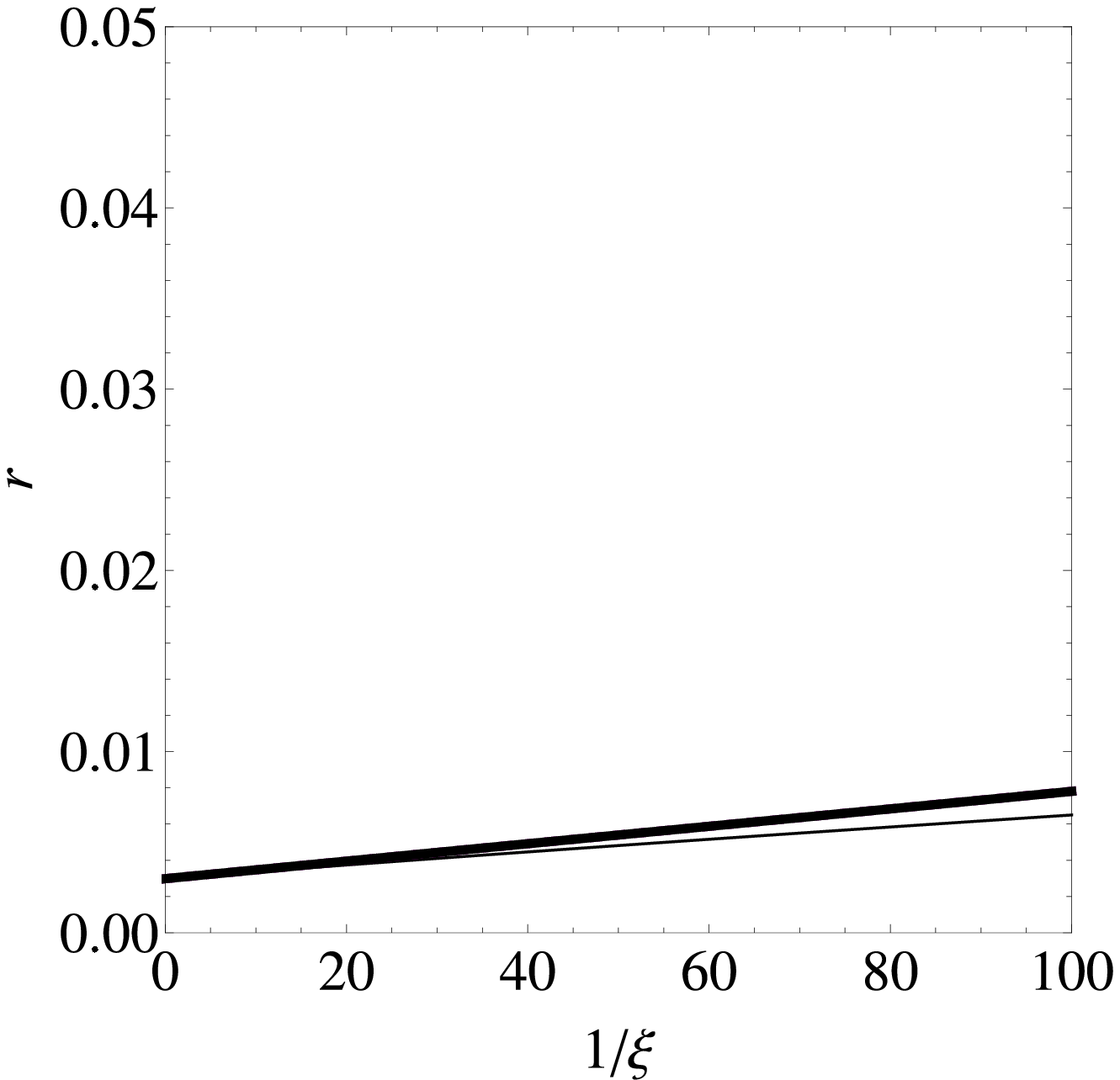}\\
  \caption{(color online). The LO, NLO, NNLO, NNNLO values of cosmological parameters $A_s$, $n_s$, $r$ in an interval of $\xi$. We set $\Lambda=0.9,0.5,0.1$ from top to bottom panels, and replace the left hand side of NNLO Eqs. \eqref{As}, \eqref{ns} with the best-fit values from Planck 2015 results \cite{Ade:2015xua}. In all panels, the thin solid lines, blue dashed lines, purple dotted lines and thick solid lines depict the results accurate to LO, NLO, NNLO, NNNLO accordingly. As indicated by the differences between purple dotted lines and thick solid lines, the NNNLO relative error is less than one percent. \emph{Warning}: The bottom panels are not trustable, because the single-field approximation breaks down for $\Lambda=0.1$ as explained in Sec. \ref{sect-2f}.}\label{fig-rAn}
\end{figure}

The above simulations are based on the NNLO Eqs. \eqref{As}, \eqref{ns}, \eqref{r}. In the following, we will set the NNLO values of $A_s$, $n_s$ to their best-fit values again, and numerically evaluate the LO, NLO, NNLO contributions to Eqs. \eqref{As}, \eqref{ns}, \eqref{r} as well as their NNNLO terms. The results are presented in Fig. \ref{fig-rAn}. As three typical subcases, from top to bottom we assign $\Lambda=0.9,0.5,0.1$ in this figure. The results converge quickly as we go to higher orders. Especially, as we have expected, the relative error induced by NNNLO terms is less than one percent.

\section{Isocurvature perturbation}\label{sect-2f}
In Sec. \ref{sect-1f}, by inserting the root of Eq. \eqref{valley} into the action, we have implicitly assumed that both background fields and their perturbations evolve along the valley. This means we have ignored isocurvature perturbation transverse to the valley. It is important to check when and how the assumption is violated. For that purpose, it is urgent to work out model \eqref{act} in the full two-field formalism, which we plan to do elsewhere. Here through numerical simulations, we would like to clarify two issues. First, starting from a slow-roll position inside the valley, the background fields roll along the valley obediently. Second, in cases with a small $\Lambda$ or equivalently a large $M_p^2\lambda/(3M^{2}\xi^{2})$, the mass of isocurvature mode can be close to the Hubble parameter.

To run the numerical algorithm, we should assign the values of model parameters, the initial values and initial velocities of fields. On the one hand, by definition of the uncalibrated e-folding number Eq. \eqref{unefolds}, the initial value of $\chi$ is dictated by
\begin{equation}\label{chistar}
\mathcal{N}=\frac{1}{8M_p^2}\chi_{\ast}^2\left(1+6\xi+\frac{2M_p^2}{M^2}\frac{\lambda}{\xi}\right),
\end{equation}
where $\mathcal{N}$ and $\chi_{\ast}$ are evaluated at Hubble crossing. Given model parameters, one can determine the initial values of $\chi$, $\phi$ by combining this relation with the equation of valley \eqref{valley}. Then the initial velocities $\dot{\chi}$, $\dot{\phi}$ can be derived from the equations of motion in slow roll. On the other hand, with $n_s$ and $A_s$ set at their best-fit values, the inflation model involves four parameters $\mathcal{N}$, $M$, $\lambda$, $\xi$ constrained by two equalities \eqref{As}, \eqref{ns}. Therefore, two more equalities are in demand. In our numerical simulations, we will set $\xi=0.01$ and $M_p^2\lambda/(3M^{2}\xi^{2})=1/9,1,9$ as three typical examples, corresponding to the tail-tips in Fig. \ref{fig-rAn}.

With all parameters fixed as above, the trajectory of fields $\chi$, $\phi$ is simulated. See the black thick line in Fig. \ref{fig-V}. It is indistinguishable from the yellow line drawn from Eq. \eqref{valley}. This demonstrates that, given appropriate initial conditions, the single-field approximation in Secs. \ref{sect-1f}, \ref{sect-data} is very good at the background level.

\begin{figure}
  \centering
  \includegraphics[width=0.22\textwidth]{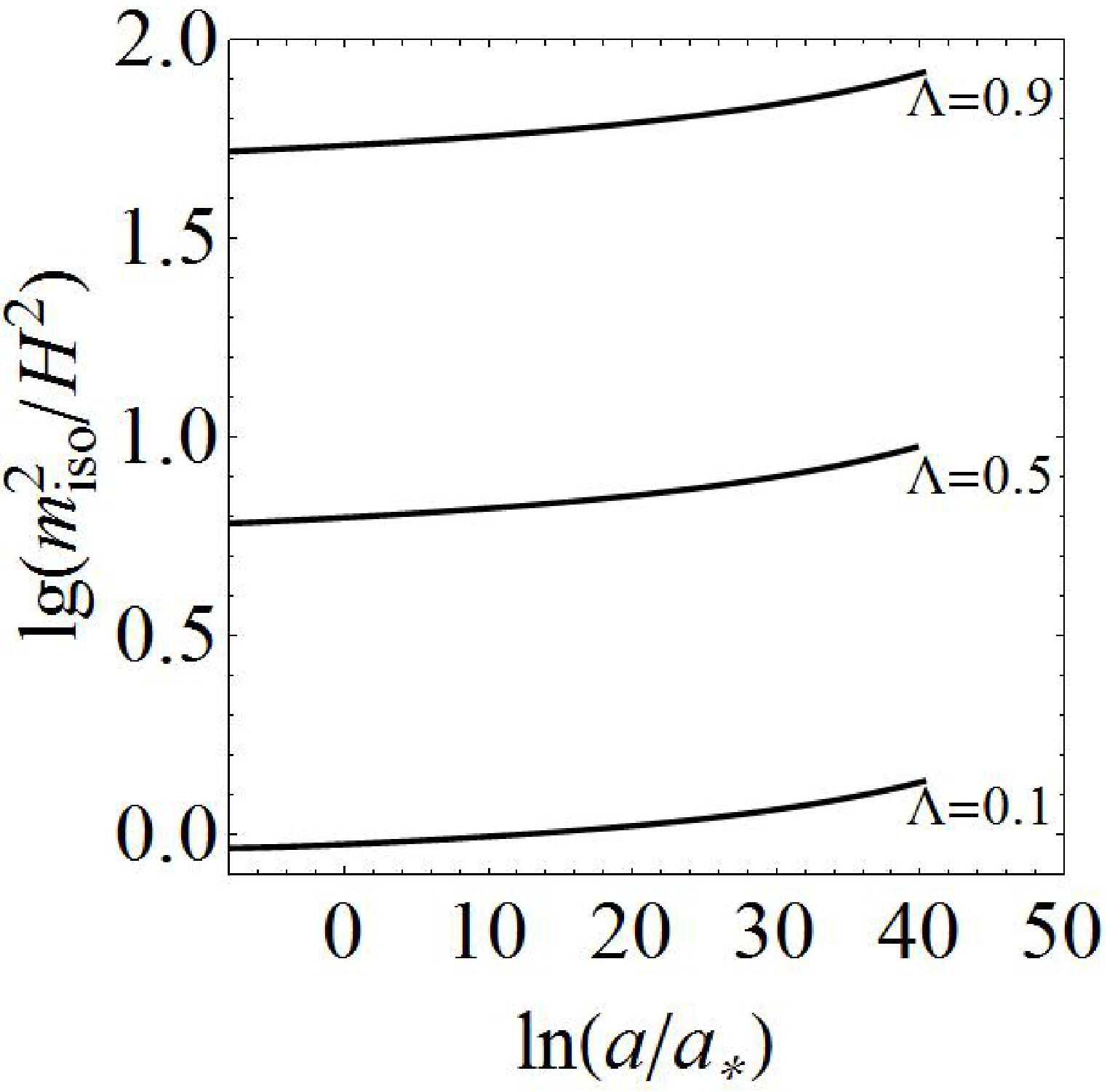}\includegraphics[width=0.22\textwidth]{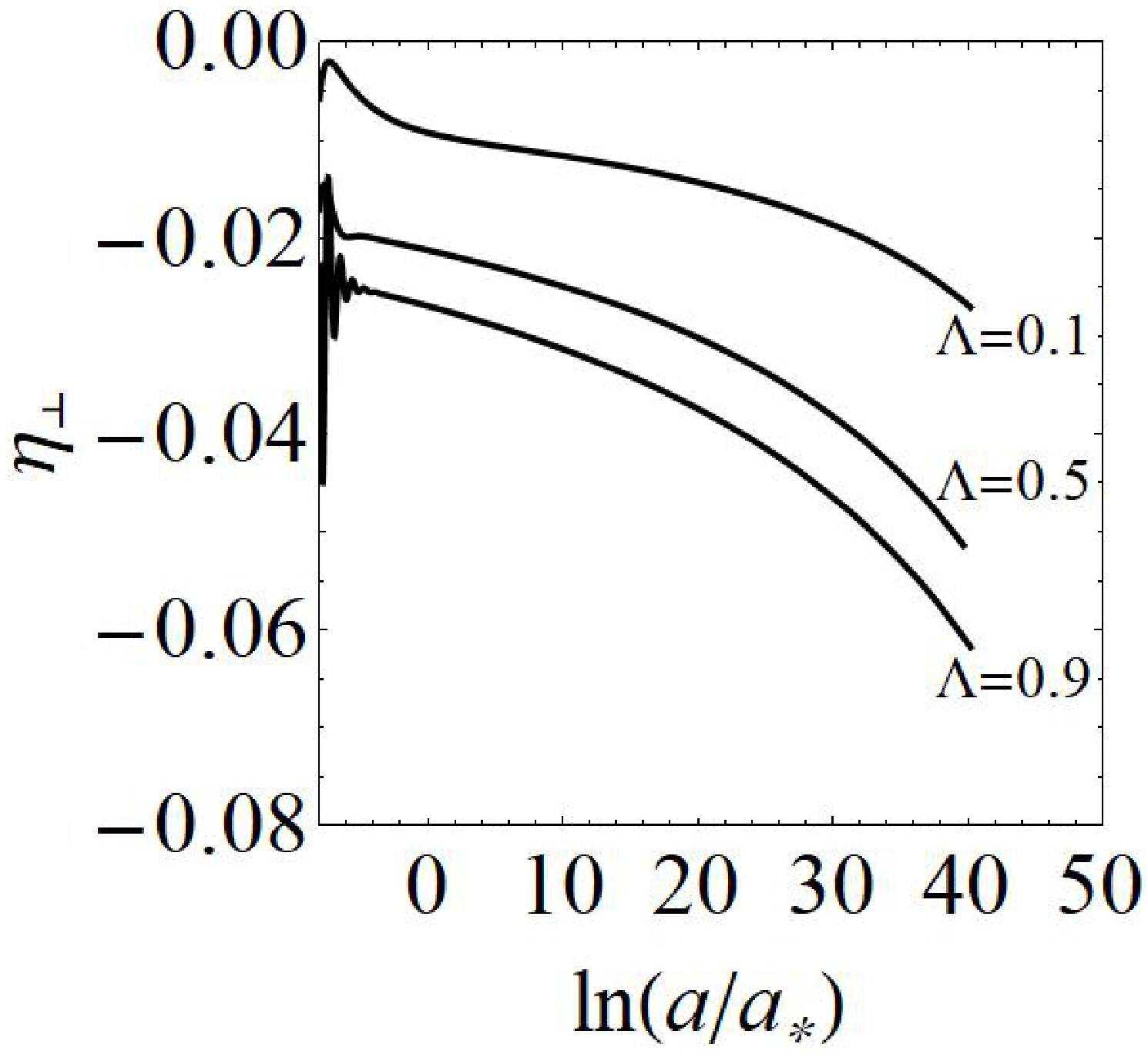}\\
  \caption{Evolution of the mass of isocurvature perturbation (left panel) and slow-turn parameter (right panel) during inflation. Initial conditions and model parameters are assigned in Sec. \ref{sect-2f}. We find $\eta_{\perp}^2\ll M^2/H^2$ in all of the three examples.}\label{fig-miso}
\end{figure}

In multi-field models with curved field spaces, the effects of isocurvature mode on the curvature power spectrum have been intensively studied in the past decade. Unfortunately, to save the space, we cannot cite hundreds of relevant works, and we have to assume that the reader is familiar with Ref. \cite{Achucarro:2010da}. In two-field inflation models, the unitary vectors tangent and normal to the trajectory, denoted by $T^{a}$ and $N^{a}$ respectively, obey Eqs. (4.1), (4.2) in Ref. \cite{Achucarro:2010da}, which mean that $T^{a}$ and $N^{a}$ turn their directions at the rate of $H\eta_{\perp}$. Roughly speaking, the inflation trajectory changes its direction by an angle $\eta_{\perp}$ within a Hubble time $1/H$. Hereafter we will refer to $\eta_{\perp}$ as the \emph{slow-turn} parameter. Following Ref. \cite{Achucarro:2010da}, we obtained analytical expressions for the mass of isocurvature mode\footnote{In Ref. \cite{Achucarro:2010da}, the mass of isocurvature mode is denoted by $M$, and the value of $\eta_{\perp}$ can be large. See Eqs. (2.18), (4.23) therein. In Ref. \cite{Chen:2012ge}, the squared mass of isocurvature mode is denoted by $M^2$ again, although it is different from $M^2$ in Ref. \cite{Achucarro:2010da} by $4\dot{\theta}_0^2$, see explanations in Ref. \cite{Pi:2012gf}. The readers can identify $M$ in these references with our notation $m_{\iso}$ when $\eta_{\perp}^2\ll M^2/H^2$. Recalling that the inflation trajectory changes its direction by an angle $\eta_{\perp}$ within a Hubble time $1/H$, we can take $R\dot{\theta}_0=\dot{\sigma}_0$, $\dot{\theta}_0/H=\eta_{\perp}$ in Ref. \cite{Chen:2012ge}.} $m_{\iso}$ and the slow-turn parameter $\eta_{\perp}$ in appendix \ref{app-miso}, which were then implemented in the numerical simulations of Fig. \ref{fig-miso}. The behaviors of $m_{\iso}$ and $\eta_{\perp}$ in this figure are consistent with Fig. \ref{fig-V}, where the valley gets deeper transversely (which implies a larger $m_{\iso}$) as $M_p^2\lambda/(3M^{2}\xi^{2})$ decreases, and the field trajectories are nearly straight (which correspond to small absolute values of $\eta_{\perp}$) in all of the black thick lines.

When $m_{\iso}\gg H$ and $|\eta_{\perp}|\lesssim1$, as occurred in the examples $\Lambda=0.9,0.5$, the isocurvature perturbation modifies the sound speed of curvature perturbation as
\begin{equation}
c_s^2=\left(1+4\eta_{\perp}^2\frac{H^2}{m_{\iso}^2}\right)^{-1}
\end{equation}
and thus induces a small modification to the curvature power spectrum \cite{Achucarro:2010da}
\begin{equation}\label{deltaA}
\frac{\delta A_s}{A_s}\sim4\eta_{\perp}^2\frac{H^2}{m_{\iso}^2}.
\end{equation}
It is less than a percent. Therefore, our results in Secs. \ref{sect-1f}, \ref{sect-data} are robust for the small $M_p^2\lambda/(3M^{2}\xi^{2})$ models although we have neglected the modification of sound speed.

However, in the case $\Lambda=0.1$, the value of $m_{\iso}$ is close to $H$, then the single-field approximation is expected to break down, and such an estimation will not be reliable. The quasi-single field inflation \cite{Chen:2009we,Chen:2009zp} would be a better scenario to study this case. To see very roughly how much $A_s$ are modified, we turn to Refs. \cite{Chen:2012ge,Pi:2012gf}. Matching to Eq. \eqref{deltaA}, we rewrite Eq. (2.5) in Ref. \cite{Chen:2012ge} as $\delta A_s/A_s\sim16\mathcal{C}\eta_{\perp}^2$. Here $\mathcal{C}$ is the same notation used in Ref. \cite{Chen:2012ge}, and Ref. \cite{Chen:2012ge} claimed that $\mathcal{C}\rightarrow H^2/(4m_{\iso}^2)$ in the large $m_{\iso}/H$ limit.\footnote{One can alternatively identify $R\dot{\theta}_0=\dot{\sigma}_0$, $\dot{\theta}_0/H=\eta_{\perp}$ in Eq. (2.5) of Ref. \cite{Chen:2012ge}, yielding $\delta A_s/A_s\sim8\mathcal{C}\eta_{\perp}^2$. This indicates that there is a slight disagreement between Ref. \cite{Achucarro:2010da} and Refs. \cite{Chen:2012ge,Pi:2012gf}, but it does not matter much to our analysis here. We take the more severe relation in our analysis.} From Figure 2 in Ref. \cite{Chen:2012ge}, we can read off $\mathcal{C}\lesssim5$ around $m_{\iso}/H\gtrsim1$. Applied to our $\Lambda=0.1$ case, it yields roughly $\delta A_s/A_s\lesssim(10\eta_{\perp})^2$, which is several percents, not as bad as we have expected.

In any case, the isocurvature perturbation generates a larger curvature power spectrum than the prediction of single-field approximation, and thus a smaller tensor-to-scalar ratio. Such an effect tends to be more significant in the small $\Lambda$ case. Recall that the single-field approximation in Secs. \ref{sect-1f} and \ref{sect-data} predict a larger $r$ for larger $\Lambda$. So we conclude that the small-$\Lambda$ or large $M_p^2\lambda/(3M^{2}\xi^{2})$ models are less interesting for observations of the tensor mode, and dirty for theoretical calculations.

As pointed out in Refs. \cite{Chen:2009we,Chen:2009zp}, in the range $m_{\iso}/H\leq3/2$, if the value of $m_{\iso}$ is smaller, the isocurvature mode will affect the curvature mode longer after Hubble crossing. The case $m_{\iso}\ll H$ is especially noteworthy, which we have not inspected so far partially because it falls into the small-$\Lambda$ models. In this case, the super-Hubble effect is most prominent, and thus the curvature power spectrum will be modulated if the end conditions of inflation are not homogeneous on different Hubble patches. It would be interesting to embed the action \eqref{act} into a more realistic model with additional scalar fields, and study such an effect in the framework of multi-brid inflation models \cite{Sasaki:2008uc,Naruko:2008sq}. In the present paper, we have implicitly assumed that the inflation ends homogeneously when our model is fully fledged.




\section{Discussion}\label{sect-disc}
Combining the Higgs inflation model and the Starobinsky's $R^2$ inflation model together, we have found that there is a cosmological attractor for inflation. It is situated in a valley of the potential landscape, deviating often from both the Higgs inflation trajectory and the Starobinsky inflation trajectory. Along the valley, slow-roll inflation takes place in the large field regime. The single-field approximation is powerful in analyzing this model and predicts $n_s$, $r$ with similar behaviors to $\alpha$-attractors. Under the single-field approximation, we studied $A_s$, $n_s$, $r$ in this model to an accuracy of one percent, and found the value of $r$ can be larger than $0.03$ in certain parametric region. We are waiting for the observational constraint on $r$ in the near future.

In two-field inflation models, the isocurvature perturbation usually works as a source term for curvature perturbation on super-horizon scales. In this paper, we restrict inflatons to the valley, or technically, by putting inflatons at a slow-roll position inside the valley initially. With initial conditions of this sort, the single-field approximation is valid unless the value of $M_p^2\lambda/(3M^{2}\xi^{2})$ is large. It would be interesting to go beyond the single-field approximation by starting with different initial conditions \cite{Kaiser:2012ak,Greenwood:2012aj,Kaiser:2013sna,Schutz:2013fua} or by exploring the zone of large $M_p^2\lambda/(3M^{2}\xi^{2})$. To avoid further complexity, we have also assumed the homogeneous end conditions of inflation. In more complicated models with inhomogeneous end conditions, one can take such an effect into account by following Refs. \cite{Sasaki:2008uc,Naruko:2008sq}, which is also interesting for future research.

In appendix \ref{app-vall}, we proposed a method to analytically locate the valley of potential in curved field spaces. It works well in the model we studied here. It would be interesting to test or improve this method in more complicated models \cite{Gorbunov:2013dqa,Myrzakulov:2015qaa,Myrzakulov:2016tsz,Brooker:2016oqa,Kannike:2015apa,Salvio:2015kka,Salvio:2016vxi}.

\vspace{0.5cm}
\noindent
{\bf Note added:}
{Soon after this paper appeared on the arXiv, Ref. \cite{Ema:2017rqn} appeared, studying also the model \eqref{act}. A comparison is compulsory here. Ref. \cite{Ema:2017rqn} is restricted to background dynamics, especially the small kinetic mixing case ($h\lesssim M_P$, or in our notations $\chi\lesssim M_p$). The theme of our paper is the valley-like landscape and the $\alpha$-attractor-like primordial perturbations. Our paper paid more attention to cosmological parameters, but overlooked the cut-off scale, the perturbativity of the quantum field theory and the cases of metastable electroweak vacuum. Ref. \cite{Ema:2017rqn} paid more attention to constraints from theory of particle physics, but omitted the cosmological perturbations, the primordial power spectra and the tensor-to-scalar ratio. Even in studying the background dynamics, the approaches are different. Ref. \cite{Ema:2017rqn} deals with a two-field system under the small kinetic mixing assumption. We transform the model into a one-field system with the help of the valley equation, and then study the system by series expansion.}

\begin{acknowledgments}
This work is supported by the National Natural Science Foundation of China (Grants No. 11105053 and No. 91536218). We are grateful to Jia-Yin Shen for enlightening discussions and the referee for pushing us to higher precision. T. W. is indebted to Shi-Ying Cai for encouragement and support.
\end{acknowledgments}

\appendix

\section{Exploring the valley}\label{app-vall}
To pave the way for studying inflation along the valley, we have to answer the question of where the valley is. This can be raised as a mathematical problem: for a 2-dimensional surface $z=V(x,y)$ in the 3-dimensional curved space
\begin{equation}\label{metric}
ds^2=h_{ab}(x,y)dx^adx^b+dz^2
\end{equation}
with $x^a=x,y$ and $a=1,2$, what is the definition or equation of valley on this surface?

In this appendix, we will try to answer the above question. We will firstly elaborate on a rudimentary method in subsection \ref{subapp-Eucl} for Euclidean space, and then present a more powerful method in \ref{subapp-curv} for curved space. Here the convention of notations is independent of that in the main text.

\subsection{Valley in Euclidean space}\label{subapp-Eucl}

Near a point ($x_0$, $y_0$, $z_0$) on the surface $z=V(x,y)$, the function $V(x,y)$ can be expanded in Taylor series
\begin{eqnarray}\label{Tay1}
\nonumber V(x,y)&=&z_0+V_{x_0}\Delta x+V_{y_0}\Delta y+\frac{1}{2}V_{x_0x_0}(\Delta x)^2\\
&&+\frac{1}{2}V_{y_0y_0}(\Delta y)^2+V_{x_0y_0}\Delta x\Delta y+\mathcal{O}(\Delta^3),
\end{eqnarray}
in which $\Delta x=x-x_0$, $\Delta y=y-y_0$. Their third and higher order terms, denoted by $\mathcal{O}(\Delta^3)$ here, are negligible. In our convention,
\begin{eqnarray}
\nonumber&&V_{x}=\frac{dV}{dx},~~~~V_{x_0}=\left.\frac{dV}{dx}\right|_{x=x_0,y=y_0},\\
&&V_{xx}=\frac{d^2V}{dx^2},~~~~V_{x_0x_0}=\left.\frac{d^2V}{dx^2}\right|_{x=x_0,y=y_0},
\end{eqnarray}
and so on. If we introduce an angle $\theta$ of the form
\begin{equation}\label{theta}
\cos\theta=\frac{V_{x_0}}{\sqrt{V_{x_0}^2+V_{y_0}^2}},~~~~\sin\theta=\frac{V_{y_0}}{\sqrt{V_{x_0}^2+V_{y_0}^2}},
\end{equation}
then locally $\Delta u=\cos\theta\Delta x+\sin\theta\Delta y$, $\Delta v=\cos\theta\Delta y-\sin\theta\Delta x$ are two orthogonal directions. After fixing ($x_0$, $y_0$), we can take $V(x,y)$ as a function of ($\Delta u$, $\Delta v$), and expand it formally in polynomials of $\Delta u$ and $\Delta v$. That is
\begin{eqnarray}\label{Tay2}
\nonumber V(x,y)&=&V(x_0+\cos\theta\Delta u-\sin\theta\Delta v,y_0+\sin\theta\Delta u+\cos\theta\Delta v)\\
\nonumber&=&z_0+V_{1}\Delta u+V_{2}\Delta v+\frac{1}{2}V_{11}(\Delta u)^2\\
&&+\frac{1}{2}V_{22}(\Delta v)^2+V_{12}\Delta u\Delta v+\mathcal{O}(\Delta^3).
\end{eqnarray}
Here the series coefficients $V_{1}=dV/d\Delta u$, $V_{2}=dV/d\Delta v$, $V_{11}=d^2V/d\Delta u^2$, and $V_{22}$, $V_{12}$ likewise, are evaluated at $\Delta u=0$, $\Delta v=0$. Explicitly, they are
\begin{eqnarray}\label{V1V2}
\nonumber V_{1}&=&V_{x_0}\cos\theta+V_{y_0}\sin\theta,\\
\nonumber V_{2}&=&V_{y_0}\cos\theta-V_{x_0}\sin\theta,\\
\nonumber V_{11}&=&V_{x_0x_0}\cos^2\theta+V_{y_0y_0}\sin^2\theta+2V_{x_0y_0}\sin\theta\cos\theta,\\
\nonumber V_{22}&=&V_{x_0x_0}\sin^2\theta+V_{y_0y_0}\cos^2\theta-2V_{x_0y_0}\sin\theta\cos\theta,\\
V_{12}&=&(V_{y_0y_0}-V_{x_0x_0})\sin\theta\cos\theta+V_{x_0y_0}(\cos^2\theta-\sin^2\theta).
\end{eqnarray}
It is easy to see that $V_{2}=0$, thus the first-order term $\Delta v$ is absent naturally in Eq. \eqref{Tay2}. This suggests that, if the point ($x_0$, $y_0$, $z_0$) is located in a valley, the valley must go along the direction $\Delta u$ locally.

Intuitively, we can define the valley as a continuum of points that locally minimize $V(x,y)$ in the transverse direction $\Delta v$. By this definition, a \emph{necessary} condition for a valley passing through the point ($x_0$, $y_0$, $z_0$) is
\begin{equation}
V_{22}>0,~~~~V_{12}=0
\end{equation}
in Eq. \eqref{Tay2}. Inserting Eqs. \eqref{theta}, \eqref{V1V2} into $V_{12}=0$, we get the equation of valley
\begin{equation}\label{vaconE}
V_{xy}(V_{x}^2-V_{y}^2)=V_{x}V_{y}(V_{xx}-V_{yy}),
\end{equation}
whose solutions ($x_0$, $y_0$) are candidates of valley. As will be clear in the next subsection, this equation is valid only for the Euclidean space.

\subsection{Valley in curved space}\label{subapp-curv}
Alternatively, in the contour map, we can define the valley as a trajectory composed of the sparsest point of each contour. According to this definition, we should minimize the gradient $h^{ab}\partial_{a}V\partial_{b}V$ subject to the the constraint $V-z=0$. Here $\partial_{a}=\partial_{x},\partial_{y}$ and $a=1,2$. This task can be accomplished by writing down the Lagrangian
\begin{equation}
\mathcal{L}(x,y,\lambda)=h^{ab}(x,y)\partial_{a}V(x,y)\partial_{b}V(x,y)+\lambda\left[V(x,y)-z\right]
\end{equation}
with the Lagrange multiplier $\lambda$, and then applying the ordinary Lagrange multiplier method,
\begin{eqnarray}
\nonumber\frac{\partial\mathcal{L}}{\partial x}&=&\partial_{x}\left(h^{ab}\partial_{a}V\partial_{b}V\right)+\lambda V_{x}=0,\\
\nonumber\frac{\partial\mathcal{L}}{\partial y}&=&\partial_{y}\left(h^{ab}\partial_{a}V\partial_{b}V\right)+\lambda V_{y}=0,\\
\frac{\partial\mathcal{L}}{\partial\lambda}&=&V(x,y)-z=0.
\end{eqnarray}
From the former two equations, we can eliminate the Lagrange multiplier $\lambda$ to get the equation of valley
\begin{equation}\label{vacon}
V_{x}\partial_{y}\left(h^{ab}\partial_{a}V\partial_{b}V\right)=V_{y}\partial_{x}\left(h^{ab}\partial_{a}V\partial_{b}V\right).
\end{equation}
This equation is a \emph{necessary} condition for valley.

Specified to the Euclidean space, $h^{xx}=h^{yy}=1$, $h^{xy}=0$, this equation reduces to Eq. \eqref{vaconE}. Therefore, we conclude that the results in subsection \ref{subapp-Eucl} are valid for the Euclidean space, but the definition and condition in the present subsection hold more generally in curved space.

As an application, Eq. \eqref{vacon} can be utilized to locate the valley in the potential landscape for model \eqref{act}. In this case, the two fields $\phi,\chi$ are identified as coordinates $x,y$, and the field space is curved as is evident from the kinetic terms in Eq. \eqref{act}. Hence the metric \eqref{metric} takes the form
\begin{equation}
ds^2=d\phi^2+e^{-\sqrt{\frac{2}{3}}\phi/M_p}d\chi^2+dz^2,
\end{equation}
and Eq.  \eqref{vacon} becomes
\begin{equation}
V_{\phi}\partial_{\chi}\left(V_{\phi}^2+e^{\sqrt{\frac{2}{3}}\phi/M_p}V_{\chi}^{2}\right)=V_{\chi}\partial_{\phi}\left(V_{\phi}^2+e^{\sqrt{\frac{2}{3}}\phi/M_p}V_{\chi}^{2}\right).
\end{equation}

\section{Analytical expressions of $m_{\iso}$ and $\eta_{\perp}$}\label{app-miso}
In Ref. \cite{Achucarro:2010da}, it was shown that the effective mass of the isocurvature mode
\begin{equation}\label{iso-mass}
m_{iso}^2=V_{NN}+H^2M_p^2\epsilon\mathbb{R},
\end{equation}
where $\epsilon=-\dot{H}/H^2$ is the slow-roll parameter, and $\mathbb{R}$ is the Ricci scalar constructed out of the field metric $h_{ab}$. For our model \eqref{act}, one has
\begin{equation}
h_{ab}=\diag\left(1,e^{-\sqrt{\frac{2}{3}}\phi/M_p}\right),~~~~\mathbb{R}=-\frac{1}{3M_p^2}.
\end{equation}
Following Ref. \cite{Achucarro:2010da}, the slow-turn parameter
\begin{equation}
\eta_{\perp}=\frac{V_{N}}{H}\left(\dot{\phi}^2+e^{-\sqrt{\frac{2}{3}}\phi/M_p}\dot{\chi}^2\right)^{-\frac{1}{2}}.
\end{equation}
In the above, $V_{N}$ and $V_{NN}$ are the covariant derivatives of $V$ with respect to the fields in the transverse direction $N^{a}$. In our model \eqref{act}, they are
\begin{eqnarray}
\nonumber V_{N}&=&N^{a}\partial_{a}V\\
\nonumber&=&\left(\dot{\phi}^2+e^{-\sqrt{\frac{2}{3}}\phi/M_p}\dot{\chi}^2\right)^{-\frac{1}{2}}\left(-e^{-\sqrt{\frac{1}{6}}\phi/M_p}\dot{\chi}V_{\phi}+e^{\sqrt{\frac{1}{6}}\phi/M_p}\dot{\phi}V_{\chi}\right),\\
\nonumber V_{NN}&=&N^{a}N^{b}\left(\partial_{a}\partial_{b}V-\Gamma^{c}_{ab}\partial_{c}V\right)\\
&=&\left(\dot{\phi}^2+e^{-\sqrt{\frac{2}{3}}\phi/M_p}\dot{\chi}^2\right)^{-1}\left[e^{-\sqrt{\frac{2}{3}}\phi/M_p}\dot{\chi}^2V_{\phi\phi}-2\dot{\phi}\dot{\chi}V_{\phi\chi}+e^{\sqrt{\frac{2}{3}}\phi/M_p}\dot{\phi}^2V_{\chi\chi}-\frac{1}{\sqrt{6}M_p}\left(2\dot{\phi}\dot{\chi}V_{\chi}+\dot{\phi}^2V_{\phi}\right)\right],
\end{eqnarray}
in which
\begin{equation}
N^{a}=\left(\dot{\phi}^2+e^{-\sqrt{\frac{2}{3}}\phi/M_p}\dot{\chi}^2\right)^{-\frac{1}{2}}\left(-e^{-\sqrt{\frac{1}{6}}\phi/M_p}\dot{\chi},e^{\sqrt{\frac{1}{6}}\phi/M_p}\dot{\phi}\right).
\end{equation}

\end{document}